\documentclass[12pt]{iopart}
\pdfoutput=1
\usepackage{color}
\usepackage{graphicx}
\usepackage{amsmath, amsthm, amssymb}
\usepackage{enumerate}
\usepackage[normalem]{ulem}
\definecolor{Gray}{gray}{0.7}
\usepackage{citesort}
\usepackage{tabularx}
\usepackage{colortbl}
\usepackage[up]{subfigure}

\newcommand{\be}{\begin{equation}}
\newcommand{\ee}{\end{equation}}
\newcommand{\bea}{\begin{eqnarray}}
\newcommand{\eea}{\end{eqnarray}}

\begin{document}

\title[Velocity distribution of driven granular gases]{Velocity distribution of driven granular gases} 
\author{V. V. Prasad$^{1,2,5}$, 
Dibyendu Das$^3$, 
Sanjib Sabhapandit$^4$ and 
R. Rajesh$^{2,5}$}

\address{$^1$ Department of Physics of Complex Systems, Weizmann Institute of Science, Rehovot 7610001, Israel}
 \address{$^2$ The Institute of Mathematical Sciences, C.I.T. Campus, Taramani, Chennai 600113, India}
\address{$^3$ Department of Physics, Indian Institute of Technology, Bombay, Powai, Mumbai-400076, India}
\address{$^4$ Raman Research Institute, Bangalore - 560080, India}
\address{$^5$ Homi Bhabha National Institute, Training School Complex, Anushakti Nagar, Mumbai 400094, India}
 \ead{prasad.vv@weizmann.ac.il, dibyendu@phy.iitb.ac.in, sanjib@rri.res.in, rrajesh@imsc.res.in}

\date{\today}

\begin{abstract}
The granular gas is a paradigm for understanding the effects of inelastic interactions in granular materials. Kinetic theory provides a 
general theoretical framework for describing the granular gas. Its central result is that the tail of the velocity distribution of a driven granular gas is a stretched exponential that,
counterintuitively, decays slower than that of the corresponding elastic gas in equilibrium. However, a 
derivation of this result starting from a microscopic model is lacking. Here, we obtain analytical results for  a 
microscopic model for a granular gas where particles with two-dimensional velocities are driven homogeneously and isotropically  by reducing the velocities by a factor and adding a stochastic noise. We find two universal regimes.  For generic physically relevant driving, we find that 
the tail of the velocity distribution is a Gaussian with additional logarithmic corrections. Thus, the velocity distribution decays faster than the corresponding equilibrium gas. The second universal regime is less generic and corresponds to the scenario described by kinetic theory. Here, the velocity distribution is shown to decay as an exponential with additional logarithmic corrections, in contradiction to the predictions 
of the phenomenological kinetic theory, necessitating a re-examination of its basic assumptions.
\end{abstract}

\maketitle

\section{\label{sec1-Introduction}Introduction}

The velocity distribution of a gas in equilibrium is well-known to be
Maxwellian (Gaussian). What is the velocity distribution for a collection of inelastic particles that is
driven to a steady state through continuous injection of energy and dissipative collisions? This is the central question
in the kinetic theory for dilute inelastic gases --- which is widely used in developing phenomenological
models for driven granular systems. Within  kinetic theory, which ignores correlations between pre-collision velocities  (molecular chaos hypothesis),
for homogeneous, isotropic  heating through a thermal bath,  the tail of the
velocity distribution  is a stretched exponential $P(v) \sim \exp(-a |v|^{\beta})$ 
with a universal exponent 
$\beta=3/2$~\cite{Noije:98}. This result is counterintuitive as it implies that larger speeds are more
probable in inelastic systems than the corresponding elastic system with the same mean energy.
A derivation of the kinetic theory result,  starting from
a microscopic model is lacking.  In addition, 
experiments and large scale simulations (see below) are unable to unambiguously
determine the tails of the distribution
and hence, a convincing answer to the question is still lacking.
In this paper, starting from a microscopic model for a driven inelastic
gas, using exact analysis we show that, for physically relevant noise distributions, $\beta=2$, albeit with additional
logarithmic corrections such that the tails of the velocity distribution decrease faster than Gaussian.  The  kinetic theory description  with thermal bath corresponds to a special limiting case of our model, for which we obtain $\beta=1$ with additional logarithmic corrections.

The tails of the velocity distribution have been studied in several experiments and
large scale computer simulations. A review of results may be found in a recent review article~\cite{windows2017granular}. Experimental systems of driven granular gases comprise of collections of granular particles such
as steel balls or glass beads that undergo  inelastic collisions and are driven either through collisions with vibrating 
walls~\cite{Clement:91,Warr:95,Kudrolli:97,Olafsen:98,Olafsen:99,Losert:99,Kudrolli:00,Rouyer:00,Blair:01,Vanzonexpt:04,reis2007forcing,wang2009particle,Scholz:17,vilquin2018shock}, 
or bilayers where only the bottom layer is vibrated~\cite{Baxter:03,baxter2007temperature,windows2013boltzmann}, 
or by application of volume forces using electric~\cite{Aranson:02,Kohlstedt:05} or magnetic fields~\cite{Schmick:08,Falcon:2013}. In addition there are experiments done in microgravity~\cite{tatsumi2009experimental,grasselli2015translational,hou2008velocity} and on different shapes like vibrated dumbbells~\cite{wildman2009granular}.
Some of the experiments observe a universal stretched exponential form with 
$\beta\approx 1.5$ for various parameters of the system~\cite{Losert:99,Rouyer:00,Aranson:02,reis2007forcing,wang2009particle,tatsumi2009experimental,Scholz:17,vilquin2018shock}, while 
other experiments find that $\beta$ differs from $3/2$ and lies between  $1$ and $2$ or is a gaussian, and may
depend on the
driving parameters~\cite{Olafsen:99,Kudrolli:00,Blair:01,Falcon:2013,grasselli2015translational,windows2013boltzmann,wildman2009granular,Schmick:08,hou2008velocity,baxter2007temperature}.
Numerical simulations~\cite{puglisi1998clustering,puglisi1999kinetic,Moon:01,Vanzon:04,Vanzon:05,Cafiero:2002,burdeau2009quasi,gayen2008orientational,gayen2011effect,rui2011velocity,Prosendas:18,kang2010granular} have also been inconclusive. The velocity distribution for a one dimensional gas driven through a thermal bath is gaussian in the quasi-elastic limit and shows deviation from the gaussian when the collisions are inelastic~\cite{puglisi1998clustering,puglisi1999kinetic}.
For a granular gas in three dimensions, driven homogeneously with a momentum conserving noise,  it was shown that 
$\beta \approx 1.5$ for large inelasticity, while $\beta$ approaches $2$ when collisions are near-elastic~\cite{Moon:01}. When the granular gas is polydispersed, a range of $\beta$ is obtained~\cite{rui2011velocity}. 
Similar study on a bounded two dimensional granular system 
find $\beta \approx 2$ for a  range of coefficient of restitution and density~\cite{Vanzon:04,Vanzon:05}, while a two dimensional
system driven through the rotational degrees of freedom find $\beta \approx 1.42$~\cite{Cafiero:2002}. Simulations of sheared granular gases find $\beta \approx 1.5$~\cite{gayen2008orientational,gayen2011effect}, while those of bilayers are consistent with $\beta =2$~\cite{burdeau2009quasi}.
Molecular dynamics simulations of a uniformly heated granular gas with solid friction find
$\beta=2$~\cite{Prosendas:18}. Models with extremal driving find intermediate power law behaviour~\cite{kang2010granular}. 
The determination of the  tails of distributions in experiments and simulations suffer from 
poor sampling of tails as well as the presence of strong crossovers
from the behaviour of the distribution at small velocities to the asymptotic behaviour at high velocities, making 
analysis difficult.

Theoretical approaches have either used  kinetic
theory, or studied simple analytically tractable models which capture the 
essential physics. 
Within kinetic theory~\cite{Brilliantov:04}, the non-linear Boltzmann equation, describing the time
evolution of the single particle velocity distribution function in the presence of a diffusion term describing driving,
is analysed. The diffusive term corresponds to a thermal bath.
The  asymptotic behaviour of the velocity distribution, obtained by linearizing the Boltzmann equation and
balancing the diffusive term with the collisional loss term, 
is characterized by $\beta=(2+\delta)/2$~\cite{ernst2003asymptotic}, where $\delta$
describes the dependence of rate of collisions
on the relative velocity $\boldsymbol{v}_{\mathrm{rel}}$ as $|\boldsymbol{v}_{\mathrm{rel}}|^\delta$. Since granular particles undergo ballistic motion ($\delta=1$) between
collisions,  one obtains $\beta=3/2$~\cite{Noije:98}.   In the quasi-elastic limit, it may be shown that $\beta=3$~\cite{benedetto1998non,barrat2002velocity,barrat2007quasi}. Subleading corrections in some cases may also be found~\cite{ernst2006rich}. The key issue in the Boltzmann equation approach is how to model driving. This issue may be addressed by studying simple particle based models in which correlations are ignored, thus mimicking  the kinetic theory description. However, since the driving mechanism is microscopic, the drawback of phenomenological modelling of the driving, inherent in the Boltzmann equation, is overcome. The study of simple particle based microscopic models have been mostly
restricted to inelastic Maxwell gases~\cite{Bobylev:00,Ben-naim:00,Baldassarri:02,Ernst:02_a,Ernst:02,Krapivsky:02,Ben-naim:02,Antal:02,Santos:03} where each pair of
particles collide at the same rate ($\delta=0$). 
The driving is of two kinds: (1) diffusive driving (random acceleration)
where a  noise $\boldsymbol{\eta}$
is added to the velocity $\boldsymbol{v}$ of a particle, i.e., $\boldsymbol{v} \to \boldsymbol{v} + \boldsymbol{\eta}$,  and (2) dissipative
driving where the driven  particle has the magnitude of its velocity reduced, in addition to receiving a kick, i.e., $\boldsymbol{v} \to -r_w \boldsymbol{v} 
+ \boldsymbol{\eta}$, where $|r_w| \leq 1$, with $r_w=-1$ corresponding to diffusive driving. Other forms of driving that have been studied include having a random coefficient of restitution~\cite{barrat2003random}, which has been argued to reproduce experimental results better, and extremal driving where
a large amount of energy is given to a single particle at a slow rate, resulting in the velocity distribution having an intermediate power law behaviour~\cite{ben2005stationary,ben2005power}.

For diffusive driving and Gaussian noise, the velocity distribution for a one-dimensional Maxwell gas ($\delta=0$)
has a universal exponential tail ($\beta=1$) independent of
the coefficient of restitution~\cite{Ernst:02,Antal:02,Santos:03}, consistent with $\delta=0$ in the kinetic theory result.
However, the noise need not always be Gaussian. For a one-dimensional Maxwell gas  with arbitrary noise statistics and dissipative driving, 
it has been shown that the tails of the velocity distribution  are non-universal and asymptotically follow the same statistics as that of the 
noise~\cite{Prasad:17}. Results for $\delta \neq 0$ have been difficult to obtain. However, for the one dimensional gas with dissipative driving, it has been possible to obtain analytical results by analysing in detail the equations satisfied by the moments~\cite{Prasad:18}. In particular, it could be shown that for $|r_w|<1$, the velocity distribution is non-universal and follows the same statistics as the noise. However, when $r_w=1$, there is a universal regime when the velocity distribution decays as an exponential with logarithmic corrections~\cite{Prasad:18}.

Diffusive driving has the drawback that it causes the velocity of the centre of mass to diffuse. 
This leads to a continuous heating up of the system, and correlations amongst the velocities grow with time~\cite{Prasad:13}. Thus, such systems do not 
reach a steady state. However,  the results that have  been derived for diffusive driving has been  
interpreted to describe a system whose reference frame is attached to the center of mass~\cite{Williams:96,Noije:98}. Therefore, a priori, it is not clear whether such theory or numerical simulations describe experimental situations where measurements are performed in the laboratory reference frame 
and the external driving is not momentum conserving~\cite{barrat2003random}.  
Dissipative driving, on the other hand, drives the system to a steady state~\cite{Prasad:13} and is closer to the experimental situation of wall-driving  (also see discussion 
after Eq.~(\ref{wall collision2}) for motivation).

In this paper, we consider a microscopic model for a two-dimensional granular gas, where  pairs of particles undergo momentum conserving, inelastic collisions at a rate proportional to $|\boldsymbol{v}_{\mathrm{rel}}|^\delta$, where  $\boldsymbol{v}_{\mathrm{rel}}$ is the relative velocity. A particle is driven dissipatively at a constant rate as described above, i.e., $\boldsymbol{v} \to -r_w \boldsymbol{v} + \boldsymbol{\eta}$, which as a special
case includes both diffusive driving ($r_w=-1$) as well as the scenario described by kinetic theory ($r_w=1$). We consider uncorrelated noise with an isotropic distribution $\Phi(\boldsymbol{\eta})$ that behaves asymptotically as $\Phi(\boldsymbol{\eta}) \sim e^{-b |\boldsymbol{\eta}|^\gamma}$ for large $|\boldsymbol{\eta}|$ [A more precise definition of the model is in \sref{sec2-The model}]. By
analysing the equations satisfied by large moments of the velocity, we determine the tails of the velocity distribution where $|\boldsymbol{v}|^2 \gg \langle |\boldsymbol{v}|^2 \rangle$. Our main results are summarised below.
For $|r_w|<1$, we obtain two regimes, both of which do not depend on $\delta$: one for $\gamma >2$ and one for $\gamma \leq 2$. For $\gamma >2$ (the noise distribution decays faster than a gaussian), we 
obtain that the tails of the velocity distribution is universal, and  $\beta=2$ with additional logarithmic corrections, 
 i.e,
\be
\ln P(\boldsymbol{v}) = - a |\boldsymbol{v}|^2 (\ln |\boldsymbol{v}|)^\tau + \ldots, 
\tau >0, ~{\rm for}~ |r_w|<1, \gamma >2.
\label{eq:main}
\ee
For $\gamma \leq 2$, the tails of the  distribution are determined only by the noise statistics, i.e.,
\be
\ln P(\boldsymbol{v}) = - a |\boldsymbol{v}|^\gamma + \ldots, 
  ~{\rm for}~ |r_w|<1, \gamma \leq 2.
\label{eq:main1}
\ee
For $r_w=1$, we obtain two regimes, which  depends on $\delta$: one for $\gamma >\gamma^*$ and another for $\gamma \leq \gamma^*$. 
For $\gamma >\gamma^*$, we 
obtain that the tails of the velocity is universal, 
i.e,
\be
\ln P(\boldsymbol{v}) = - a |\boldsymbol{v}|^\beta (\ln |\boldsymbol{v}|)^\theta + \ldots, 
~{\rm for}~ r_w=1, \gamma > \gamma^*=\frac{2 + \min(\delta,0)}{2},
\label{eq:main2}
\ee
where
\begin{align}
&\beta=  \frac{2 + \delta}{2}, ~ \theta=0, &\delta \leq 0,\\
&\beta=  1,  \qquad \theta=\frac{\gamma}{\gamma-1}, &\delta > 0.
\end{align}
For $\gamma \leq \gamma^*$, the tails of the velocity distribution are determined only by the noise statistics, i.e.,
\be
\ln P(\boldsymbol{v}) = - a |\boldsymbol{v}|^\gamma + \ldots, 
  ~{\rm for}~  r_w=1, \gamma \leq \gamma^*=\frac{2 + \min(\delta,0)}{2}.
\label{eq:main3}
\ee
We argue
that  physically realistic noise distributions fall off faster than a Gaussian, and hence $P({\boldsymbol{v}})$ is generically
as described in \eref{eq:main}. For $r_w=1$ and $\delta=1$, we obtain $\beta=1$, in contradiction to the results from kinetic theory.

The remainder of the paper is organised as follows. We define the model precisely in \sref{sec2-The model}, along with both the motivations as well as the connections to kinetic theory. In \sref{sec3-Existence of steady state}, existence of steady state is shown analytically for the case $\delta=0$ by solving for the two point correlations. For other $\delta$, a numerical study of the temporal evolution of the energy is done.  In \sref{sec4-Moment analysis}, we do a detailed analysis of the equations satisfied by the moments of the velocity, by making an ansatz for the velocity distribution and looking for self-consistent solutions. This allows us to determine the asymptotic behaviour of the velocity distribution.  The comparison of the 
analytical results with Monte Carlo simulations and an earlier experiment is described in sections \ref{sec5-Monte-Carlo results} and \ref{sec7-Comparison with exp data} respectively. Finally in  \sref{sec8-Discussions}, we conclude 
by summarising our results and discussing their implications.

\section{\label{sec2-The model}The model}

Consider a system of $N$  identical particles labelled by $i=1, \ldots, N$, having two dimensional velocities ${\boldsymbol{v_i}}=(v_{xi},v_{yi})$.
Particles $i$ and $j$ undergo momentum conserving inelastic collision at a rate $\frac{2 \lambda_c}{N-1} |{\boldsymbol{v_i}} - {\boldsymbol{v_j}}|^\delta$,
and the new velocities ${\boldsymbol{v'_i}}$ and ${\boldsymbol{v'_j}}$ are given by
\bea
{\boldsymbol{v'_i}}={\boldsymbol{v_i}}-\alpha \left[({\boldsymbol{v_i}}-{\boldsymbol{v_j}})\cdot\hat{\boldsymbol \sigma}\right]\hat{\boldsymbol \sigma},\nonumber\\
{\boldsymbol{v'_j}}={\boldsymbol{v_j}}+\alpha \left[({\boldsymbol{v_i}}-{\boldsymbol{v_j}})\cdot\hat{\boldsymbol \sigma}\right]\hat{\boldsymbol \sigma},
\label{inelastic collision_2d}
\eea
where $\alpha=(1+r)/2$, $r$ being the coefficient of restitution, and 
$\hat{\boldsymbol \sigma}$ is a unit vector along the line joining the
centres of the particles at contact. 
We assume that $\hat{\boldsymbol \sigma}$ is randomly oriented, such that it takes a value uniformly from $[0,2 \pi)$ for each collision. Note that we have assumed
a well-mixed system, as  is also assumed in kinetic theory, such that spatial information is ignored.
Since $r\in[0,1)$, we obtain $\alpha \in[1/2,1)$. A particle $i$ is driven
at rate $\lambda_d$ and the new velocity ${\boldsymbol{v'_i}}$ is given by~\cite{Prasad:13}
\bea
{\boldsymbol{v'_i}}=-r_w {\boldsymbol{v_i}} + {\boldsymbol{\eta}}, \quad |r_w| \leq 1 ,
\label{wall collision2}
\eea 
where $r_w$ is a parameter by which the speed is decreased. The noise ${\boldsymbol{\eta}}$ is uncorrelated in time,
and drawn from a fixed distribution $\Phi(\boldsymbol{\eta})$. Note that
diffusive driving may be realized by setting $r_w=-1$ in Eq.~\eref{wall collision2}. Also, the limit $r_w=1$ may be argued to correspond to the scenario described by kinetic theory (see discussion below).

We characterize the isotropic noise distribution $\Phi({\boldsymbol{\eta}})$
by its asymptotic behaviour 
\be
\Phi({\boldsymbol{\eta}}) \sim e^{-b |{\boldsymbol{\eta}}|^\gamma}, ~b,\gamma>0,~\quad |{\boldsymbol{\eta}}| \gg \sigma_\eta,
\label{ansatz noise pdf}
\ee 
where $\sigma_\eta^2$ is the second moment. 
It is not  necessary that $\boldsymbol{\eta}$ is a Gaussian with $\gamma=2$, since the noise in a granular system is not generated from
a sum of many small stochastic events. Therefore, we keep $\gamma$ arbitrary. We have assumed a stretched exponential decay for the
noise distribution. As will turn out from the analysis, slower decay like power laws may be absorbed into $\gamma=0$ and faster decays
than stretched exponential may be absorbed into $\gamma=\infty$.

There are certain motivations for choosing the driving as in Eq.~(\ref{wall collision2}). First is that for $r_w \neq -1$, the system is driven to a steady state (see \sref{sec3-Existence of steady state}), overcoming the drawbacks of diffusive driving for which there is no steady state. Second,  the limit $r_w=1$ is the scenario described by kinetic theory and hence provides a more rigorous check for its predictions. This may be argued as follows.
Let $P(\boldsymbol{v},t)$ denote the probability that a randomly chosen particle has velocity $\boldsymbol{v}$ at time $t$. Its time evolution is described by the
master equation:
\begin{align}
\frac{d P(\boldsymbol{v},t)}{dt} &=
 \lambda_c \int\!\!\int\!\! \int\! d\hat{\boldsymbol \sigma}  d\boldsymbol{v}_1  d\boldsymbol{v}_2 |\boldsymbol{v}_1-\boldsymbol{v}_2|^\delta P(\boldsymbol{v}_1,t) P(\boldsymbol{v}_2,t) 
 \delta\left({\boldsymbol{v_1}}-\alpha \left[({\boldsymbol{v_1}}-{\boldsymbol{v_2}})\cdot\hat{\boldsymbol \sigma}\right]\hat{\boldsymbol \sigma} -\boldsymbol{v}\right) \nonumber \\
&- 2 \lambda_c \int  d\boldsymbol{v}_2 |\boldsymbol{v}-\boldsymbol{v}_2|^\delta P(\boldsymbol{v},t) P(\boldsymbol{v}_2,t) 
-\lambda_d P(\boldsymbol{v},t) \nonumber \\
&+\lambda_d \int \int d\boldsymbol{\eta} d\boldsymbol{v}_1 \Phi(\boldsymbol\eta) P(\boldsymbol{v}_1,t) \delta\left[ -r_w \boldsymbol{v}_1+\boldsymbol{\eta} -\boldsymbol{v} \right],
\label{eq:evolution}
\end{align}
where we have used product measure for  the joint distribution $P(\boldsymbol{v}_1,\boldsymbol{v}_2)= P(\boldsymbol{v}_1) P(\boldsymbol{v}_2)$ due to lack of correlations between velocities of different
particles, arising from the fact that pairs of particles
collide at random (see also \sref{sec3-Existence of steady state}, where the two-point correlations are shown to vanish for $\delta=0$). The first two terms on the right hand side of Eq.~\eref{eq:evolution}
describe the gain and loss terms due to inter-particle collisions. The third and fourth terms on the right hand side describe the loss and
gain terms due to driving. The driving terms may be analysed for small $|{\boldsymbol\eta}|$ as follows. Let
\be
I_D= -\lambda_d P(\boldsymbol{v},t)
+\lambda_d \int \int d\boldsymbol{\eta} d\boldsymbol{v}_1 \Phi(\boldsymbol\eta) P(\boldsymbol{v}_1,t) \frac{1}{r_w}\delta\left[\frac{\boldsymbol{\eta} -\boldsymbol{v}}{r_w}- \boldsymbol{v}_1 \right].
\ee
Integrating over $\boldsymbol{v}_1$, and using the symmetry property $P(\boldsymbol{v}) = P(-\boldsymbol{v})$, we obtain
\be
I_D= -\lambda_d P(\boldsymbol{v},t)
+\frac{\lambda_d}{r_w} \int \int d\boldsymbol{\eta} d\boldsymbol{v} \Phi(\boldsymbol\eta) P\left(\frac{\boldsymbol{v} -\boldsymbol{\eta}}{r_w},t \right).
\label{eq:aaaaa}
\ee
Setting $r_w=1$, and Taylor expanding the integrand about $|{\boldsymbol{\eta}}|=0$, and then integrating over $\boldsymbol{\eta}$, Eq.~(\ref{eq:aaaaa}) reduces to
\be
I_D= \frac{\lambda_d \langle |\boldsymbol{\eta}|^2 \rangle}{2} \nabla^2 P(\boldsymbol{v}) +\mathrm{ higher ~order ~terms}, \quad r_w=1.
\label{eq:aaaaaa}
\ee
When the higher order terms are ignored, the resulting equation for $P(\boldsymbol{v})$ for $r_w=1$ is the same as that was analysed in Ref.~\cite{Noije:98} to obtain the well-known result of $\ln P(\boldsymbol{v}) \sim -|\boldsymbol{v}|^{3/2}$. It is not apriori clear whether this truncation is valid, as the tails of the velocity distribution could be affected by tails of the noise distribution, in which case higher order moments of noise may contribute.

Third, dissipative driving may be motivated by modelling the collisions of  a particle with a massive wall. \Eref{wall collision2} may be derived by defining the particle-wall coefficient of restitution to be $r_w$, and  assuming that the wall is massive compared to the particles, and also that the collision times are random~\cite{Prasad:13}.
Within this motivation, $r_w$ is positive, and one may also argue that for physically relevant noise distributions $\gamma \gg 2$, since the noise is often bounded from above.  For example, for a sinusoidally oscillating
wall, if the collision times are assumed to be random, then it is straightforward to show that 
$\Phi(\eta) \sim (c^2-\eta^2)^{-1/2}$, with $\eta\in (-c,c)$,  corresponding to $\gamma=\infty$. This analogy of dissipative driving  with wall-collisions is strictly valid only in one dimension, as it assumes that the wall moves colinearly with the particle velocity. For a two dimensional gas, one would expect that only the component of velocity perpendicular to the motion of the wall is reversed. A realistic model would be one where only one of the two components of velocity is reversed when driven, making the noise anisotropic. However, the assumption of isotropic noise, as assumed in this paper as well as in kinetic theory, makes calculations easier.  We note that driving only one component dissipatively will still result in a steady state, overcoming the drawbacks of diffusive driving, as the momentum in the other direction is strictly conserved. We also expect  that the results we derive for isotropic noise continue to hold for anisotropic driving, and we confirm this through detailed Monte Carlo simulations (see \sref{sec5-Monte-Carlo results}).

\section{\label{sec3-Existence of steady state} Existence of steady state}

 We first show that the system reaches a steady state when $r_w\neq -1$. The equations obeyed
by the set of two-point correlation functions close and may be solved explicitly when $\delta=0$.
We follow closely the
method of calculation used for determining the same for the one-dimensional Maxwell gas~\cite{Prasad:13}.

Let ${\boldsymbol v}= (v_x, v_y)$. We are interested in the evolution of the following two-point correlation functions
\begin{align}
&\Sigma^x_0(t)=\frac{1}{N}\displaystyle\sum_{i}\langle v_{ix}(t)v_{ix}(t)\rangle, &\Sigma^x_{12}(t)=\frac{1}{N(N-1)}\displaystyle\sum_{i\not=j}\langle\ v_{ix}(t)v_{jx}(t)\rangle,\nonumber\\
&\Sigma^y_0(t)=\frac{1}{N}\displaystyle\sum_{i}\langle v_{iy}(t)v_{iy}(t)\rangle, &\Sigma^y_{12}(t)=\frac{1}{N(N-1)}\displaystyle\sum_{i\not=j}\langle v_{iy}(t)v_{jy}(t)\rangle, \label{2 point correlations}\\
&\Sigma^{xy}_0(t)=\frac{1}{N}\displaystyle\sum_{i}\langle v_{ix}(t)v_{iy}(t)\rangle, &\Sigma^{xy}_{12}(t)=\frac{1}{N(N-1)}\displaystyle\sum_{i\not=j}\langle v_{ix}(t)v_{jy}(t)\rangle. \nonumber
\end{align}
 From the dynamics, as described in Eqs.~(\ref{inelastic collision_2d}) and (\ref{wall collision2}),  the exact evolution of the two-point functions may  be derived
as a set of coupled equations,  which may be written in a compact form as
\be
\frac{d{\boldsymbol{\Sigma}}(t)}{dt}={\boldsymbol{R}}{\boldsymbol{\Sigma}}(t)+{\boldsymbol{C}}.
\label{evolution equation}
\ee
Here,  the column vectors, ${\boldsymbol{\Sigma}}^T,~{\boldsymbol{C}}^T$ are given by:
\bea
{\boldsymbol{\Sigma}}(t)&=&[\Sigma^x_0(t),~\Sigma^x_{12}(t),~\Sigma^y_0(t),~\Sigma^y_{12}(t),~\Sigma^{xy}_0(t),~\Sigma^{xy}_{12}(t)]^T,\\
{\boldsymbol{C}}&=& \left[\frac{\sigma^2_\eta}{2N}, ~0, ~\frac{\sigma^2_\eta}{2N},~0,~0,~0 \right]^T,
\eea
and ${\boldsymbol{R}}$ is the matrix
{\footnotesize{
\begin{equation}
\left[\begin{array}{cccccc} 
 -A_1- A_2(1-r_w)& A_1&A_3&-A_3&0&0\\
 \frac{A_1}{N-1}& \frac{-A_1}{N-1}-2A_2 &\frac{-A_3}{(N-1)}&\frac{A_3}{(N-1)}&0&0\\
A_3&-A_3& -A_1-A_2(1-r_w)&A_1&0&0\\
\frac{-A_3}{N-1}&\frac{A_3}{N-1}&\frac{A_1}{(N-1)}&\frac{-A_1}{N-1}-2A_2&0&0\\
0&0&0&0&-A_4-A_2&A_4\\
0&0&0&0&\frac{A_4}{N-1}& \frac{-A_4}{N-1}-2A_2
\end{array}\right].
\end{equation}}}
The constants  $\{A_i\}$'s  are functions of the rates $\lambda_c$, $\lambda_d$  as well as the coefficient of restitution, $\alpha=(1+r)/2$ and $r_w$:
\begin{align}
&A_1=\lambda_c \left( 2\alpha-\frac{3\alpha^2}{2} \right),  &A_2=\lambda_d(1+r_w),\\
&A_3=\frac{\lambda_c\alpha^2}{2},  &A_4=\lambda_c(2\alpha-\alpha^2).
\end{align}

In the steady state the left hand side of Eq.~\eref{evolution equation} equals zero. Solving
the resulting linear equation, we obtain the steady state values of the different correlation functions as
\be
\Sigma_0=\frac{\lambda_d\sigma^2_\eta}{2\left[2\alpha(1-\alpha)\lambda_c+(1-r_w^2)\lambda_d\right]}+\frac{\alpha^2(1-\alpha)^2\lambda_c^2\sigma^2_\eta}{(1+r_w)[2\alpha(1-\alpha)\lambda_c+(1-r_w^2)\lambda_d]^2}\frac{1}{N}
+O\left(N^{-2}\right),
\label{SI-steady energy}
\ee
where $\Sigma_0\equiv\Sigma^x_0=\Sigma^y_0$,
and 
\be
\Sigma^x_{12}=\Sigma^y_{12}=\frac{\alpha(1-\alpha)\lambda_c \sigma^2_\eta}{2(1+r_w)[2\alpha(1-\alpha)\lambda_c+(1-r_w^2)\lambda_d]}\frac{1}{N}
+O\left(N^{-2}\right).
\label{SI-steady correlation}
\ee
When $r_w\not=-1$,  the limit $N\to \infty$  is well defined with finite non-zero value for $\Sigma_0$ [Eq.~\eref{SI-steady energy}]. On the other hand,  the correlations $\Sigma_{12}$ [Eq.~\eref{SI-steady correlation}] vanishes as $O(1/N)$.  However, when $r_w=-1$, the $O\left(N^{-1}\right)$ blows up,
and $\Sigma_0^{x,y}=\infty$ implying the 
absence of steady state. Note that, this is not the case when $r_w=1$ for which  $\Sigma_0$ has a  finite value.

The analytical calculation cannot be fully  extended to a general collision kernel where the collision rate  depends on the relative velocities ($\delta\neq 0$). However, as we show below, the two-point correlations may be expressed in terms of the mean energy of the system. Let 
$\mathbf{P}=\sum_i \boldsymbol{v}_i$ be the momentum of the centre of mass of the system. In a collision, $\mathbf{P}$ is conserved. During driving $\mathbf{P}$ changes stochastically according to
\be
\label{eq:totalmomentum}
\mathbf{P}(t+dt)=\begin{cases}
\mathbf{P}(t) - \boldsymbol{v}_i +(-r_w \boldsymbol{v}_i + \boldsymbol{\eta}), &  \mathrm{probability}=\lambda_d dt,\\
\mathbf{P}(t)  &  \mathrm{probability}=1- N \lambda_d dt.
\end{cases}
\ee
It is then straightforward to obtain that
\be
\frac{d \langle \mathbf{P}^2 \rangle}{dt}=N \lambda_d\left[ (1+r_w)^2 (\Sigma_0^x + \Sigma_0^y)+ \sigma_{\eta}^2- \frac{2 (1+r_w)}{N} \sum_i \langle \mathbf{P} \cdot \boldsymbol{v}_i   \rangle \right].
\label{eq:cccd}
\ee
In the steady state, when the left hand side of Eq.~(\ref{eq:cccd}) is equal to zero, and substituting for $\mathbf{P} = \sum _i \boldsymbol{v}_i$, we obtain
\be
\Sigma_{12}^x + \Sigma_{12}^y= \frac{\sigma_\eta^2-(1-r_w^2) (\Sigma_0^x + \Sigma_0^y)}{2(1+r_w) (N-1)}.
\label{general}
\ee
It may easily be checked that the exact solution for the case $\delta=0$ as given in Eqs.~(\ref{SI-steady energy}) and (\ref{SI-steady correlation}) satisfies Eq.~(\ref{general}). It follows from Eq.~(\ref{general}) that for all $\delta$, the correlations between the velocities of two different particles are $\mathcal{O}(N^{-1})$, and are equal to  zero in the thermodynamic limit.

\begin{figure}
\begin{center}
 \includegraphics[width=1\textwidth]{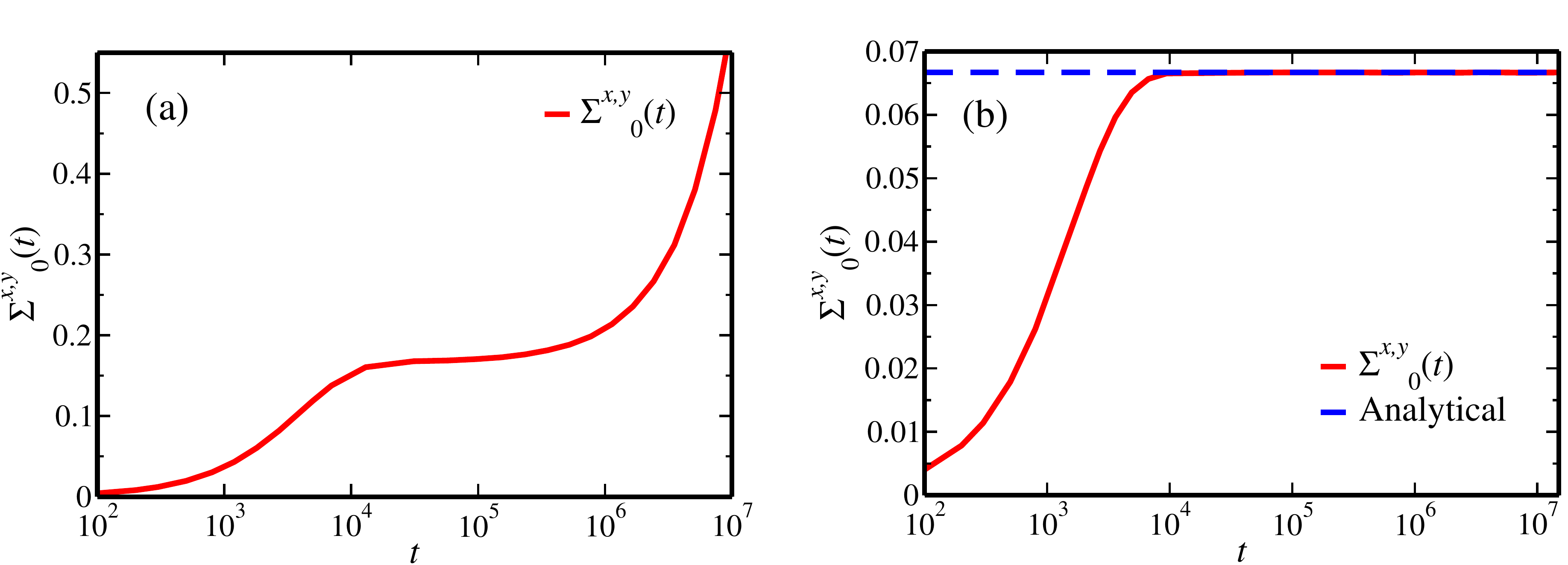}
 \end{center}
\caption{ \label{SI-f1}The time evolution of the variance $\Sigma^{x,y}_0(t)$ for a Maxwell gas  ($\delta=0$) of $N=1000$ particles and $r=0$. 
Note that the x-axis is logarithmic. Plot (a)  shows the evolution when the driving is diffusive ($r_w=-1$). 
The monotonic increase in the   variance in the large 
time limit illustrates the  lack of steady state for the system. Plot (b) shows the data for dissipative driving with $r_w=1/2$. The system 
reaches a stationary state. The dotted lines denotes the analytically obtained steady state value [see Eq.~\eref{SI-steady energy}]. In both the cases, the noise $\eta$  is drawn from a 
uniform  distribution with $\sigma^2_\eta=1/12$ and $\lambda_c=\lambda_d=1/2$.}
 \end{figure}

It is, however, not possible to determine exactly the steady state energy when $\delta\neq 0$. Nonetheless, it is possible to perform a Monte Carlo simulation for such systems.
To benchmark our simulations, we first verify the results for $\delta=0$. In  Fig.~\ref{SI-f1},  the time evolution of $\Sigma_0$, as obtained from Monte Carlo simulations, is shown
for the case $\delta=0$. 
For diffusive driving  ($r_w=-1$), shown in  Fig.~\ref{SI-f1}(a), we find
that  for large times the variance $\Sigma_0$ does not saturate but increases monotonically as a function of 
time, showing that the system does not have a steady state as shown by the analytical calculation. The lack of 
steady state is caused by the diffusion of the centre of mass due to the additive noise in the driving  that do not 
conserve the total momentum. For $r_w\not=-1$ the system reaches a steady state, as seen in Fig.~\ref{SI-f1}(b),
which shows the time evolution of $\Sigma_0$ for the case $r_w=1/2$. The numerically obtained steady state value
coincides with the analytically obtained value.

\begin{figure}
\begin{center}
 \includegraphics[width=1\textwidth]{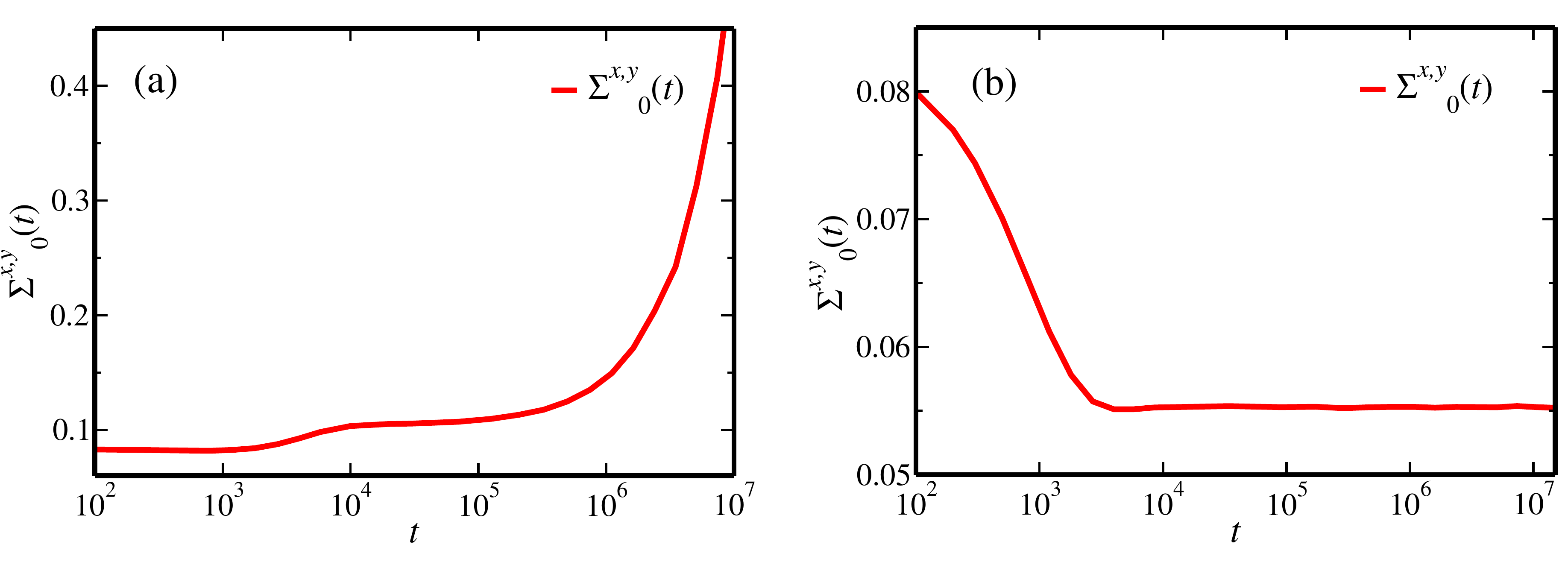}
 \end{center}
\caption{\label{SI-f2}The time evolution of the variance $\Sigma^{x,y}_0(t)$ for a granular gas ($\delta=1$) of $N=1000$ particles and $r=0$.
Note that the x-axis is logarithmic. Plot~(a) shows the evolution when the driving is diffusive ($r_w=-1$). The monotonic increase in the   variance in the large 
time limit illustrates the  lack of steady state for the system. Plot~(b) show the data for dissipative driving with $r_w=1/2$. The system 
reaches a stationary state.  In both the cases, the noise $\eta$  is drawn from a 
uniform  distribution with $\sigma^2_\eta=1/12$ and $\lambda_c=\lambda_d=1/2$.
}   
\end{figure}

Figure~\ref{SI-f2} shows the time evolution of the mean energy of
a system with $\delta=1$ (ballistic gas). The results are qualitatively the same as that obtained for the case $\delta=0$. For
diffusive driving ($r_w=-1$) the system does not reach a steady state [see Fig.~\ref{SI-f2}(a)]. However, for $r_w \not=-1$ the system reaches a steady state
as seen in  Fig. \ref{SI-f2}(b).

\section{\label{sec4-Moment analysis}Moment analysis}

The tails of the velocity distribution may be inferred  by knowing the large moments of the velocity.
The equations obeyed by the moments are obtained by multiplying Eq.~\eref{eq:evolution} by $v_x^{2 n}$ and integrating
over all velocities, or directly from Eqs.~(\ref{inelastic collision_2d}) and (\ref{wall collision2}). 
In the steady state, after setting time derivatives to zero, we obtain
\be
\label{moment equation00}
2 \lambda_c \langle |\boldsymbol{v}-\boldsymbol{v}'|^\delta v_x^{2 n}\rangle + \lambda_d (1-r_w^{2n}) \langle v_x^{2 n}\rangle =
2 \lambda_c \sum_{m=0}^{2n} t_m
+\lambda_d \sum_{m=0}^{n-1}\binom{2n}{2m} r_w^{2m} \langle v_x^{2 m}\rangle \mathcal{N}_{2n-2m},
\ee
where $\mathcal{N}_{2m}=\int d\boldsymbol{\eta}  \Phi(\boldsymbol{\eta}) \eta_x^{2m}$, the averages $\langle \cdots  \rangle$ are over the distribution $P(\boldsymbol{v}) P (\boldsymbol{v}')$, where $\boldsymbol{v}$ and $\boldsymbol{v}'$ are the velocities of two different particles, and
\be
\label{moment equation-no driving-typical term}
t_m = \int_0^{2 \pi}  \frac{d\theta}{2 \pi} \binom{2n}{m}  \langle |\boldsymbol{v}-\boldsymbol{v}'|^\delta
(a_1 v_x+a_2 v_y)^{2 n-m} (a_3 v'_x+a_4 v'_y)^{m}\rangle,
\ee
where $a_1=1-\alpha\cos^2\theta$,
$a_2=-\alpha\cos\theta\sin\theta$, 
$a_3=1-a_1$, and $a_4=-a_2$. The first sum in the right hand side of Eq.~(\ref{moment equation00}) has its origin
in inter-particles collisions while the second sum arises from driving. For large $n$, the left hand side of Eq.~\eref{moment equation00} is dominated by the first term if $\delta \geq 0$ or $r_w=1$, else it is dominated by the second term. Thus, we may write Eq.~\eref{moment equation00} as
\be
\label{moment equation}
 \langle |\boldsymbol{v}-\boldsymbol{v}'|^{\tilde{\delta}} v_x^{2 n}\rangle \sim 
 2 \lambda_c \sum_{m=0}^{2n} t_m
+\lambda_d \sum_{m=0}^{n-1}\binom{2n}{2m} r_w^{2m} \langle v_x^{2 m}\rangle \mathcal{N}_{2n-2m},~~n \gg 1
\ee
where  
\begin{equation}
\tilde{\delta}=\begin{cases}
\max(\delta,0),& |r_w|<1, \\
\delta,& r_w=1,
\end{cases}
\end{equation}
and $x\sim y$ means that $x/y = \mathcal{O}(1)$.

Since the driving is isotropic, the velocity distribution is also isotropic and hence is a function of 
only the modulus of velocity. Thus, we write
\begin{align}
P(\boldsymbol{v}) \sim e^{-a|\boldsymbol{v}|^\beta+\Psi(|\boldsymbol{v}|)},~a,\beta>0,|\boldsymbol{v}|^2 \gg \langle |\boldsymbol{v}|^2 \rangle,
\label{velocity-isotropic pdf}
\end{align} 
where the correction term is such that $|\boldsymbol{v}|^{-\beta} \Psi (|\boldsymbol{v}|) \to 0$.  The moments $M_{2n}\equiv\langle v_x^{2n}\rangle=\langle v_y^{2n}\rangle$ for this distribution
may be determined for $n \gg 1$ using a saddle point approximation (the details are given in \ref{appendix1}):
\begin{align}
&M_{2n}\sim\left[\frac{2n}{ae\beta}\right]^{\frac{2n}{\beta}}n^{\frac{2-\beta}{\beta}} e^{\Psi[(\frac{2 n}{a \beta})^{\frac{1}{\beta}}]}, 
\label{velocity-moment}\\
&\langle (a_i v_x+a_j v_y)^{2 n} \rangle \sim M_{2n} (a_i^2+a_j^2)^n,  \quad n,~m \gg 1, \label{velocity-isotropic-moment}\\
&  \langle |\boldsymbol{v}-\boldsymbol{v}'|^\delta
(a_1 v_x+a_2 v_y)^{2 n-2 m} (a_3 v'_x+a_4 v'_y)^{2 m}\rangle \sim n^{\frac{\delta}{\beta}} M_{2 n -2 m} M_{2m} (a_1^2 \!+\! a_2^2)^{n-m} (a_3^2 \!+ \!a_4^2)^{m}.
\label{velocity-coupled-moment1}
\end{align} 
The asymptotic behaviour of the moments of noise may be obtained from Eq.~(\ref{velocity-moment}) by replacing $\beta$ with $\gamma$, and $a$ by $b$, i.e.,
\be
\mathcal{N}_{2n}\equiv \langle \eta_x^{2 n}\rangle \sim\left[\frac{2n}{b e\gamma}\right]^{\frac{2n}{\gamma}}n^{\frac{2-\gamma}{\gamma}}, \quad n \gg 1.
\label{noise-moment}
\ee

It is clear that the  moments in Eqs.~\eref{velocity-moment}-\eref{noise-moment}, and hence the terms in Eq.~\eref{moment equation}, diverge factorially with $n$.
Therefore, the sums in the right hand side of Eq.~\eref{moment equation} may be approximated by the  largest terms with negligible error. The largest term could be part of the first sum or the second sum. Since, we do not apriori know which sum it belongs to, we consider both possibilities. Assuming that 
the largest term is part of the first sum, we solve the equation
\be
\label{moment-equation-coll}
 \langle |\boldsymbol{v}-\boldsymbol{v}'|^{\tilde{\delta}} v_x^{2 n}\rangle \sim \sum_{m=0}^{2 n} t_m,
\ee
while assuming that it is part of the second sum, we solve the equation
\be
\label{moment-equation-drive}
\langle |\boldsymbol{v}-\boldsymbol{v}'|^{\tilde{\delta}} v_x^{2 n}\rangle \sim  \sum_{m=0}^{n-1}\binom{2n}{2m} r_w^{2m} M_{2m}\mathcal{N}_{2n-2m}.
\ee
We denote the solution obtained for $\beta$ by solving Eq.~\eref{moment-equation-coll}, with the ansatz for the velocity distribution as in Eq.~\eref{velocity-isotropic pdf} by $\beta_c$, and that obtained by solving Eq.~\eref{moment-equation-drive} by $\beta_d$, 
where the subscripts $c$ and $d$ denote collision and driving respectively. Clearly,
\be
\beta=\min(\beta_c, \beta_d).
\label{eq:equality}
\ee

We first consider the case when driving is dominant and evaluate  $\beta_d$ by searching for self consistent solutions of Eq.~(\ref{moment-equation-drive}). This calculation is identical to that for the one dimensional granular gas because Eq.~(\ref{moment-equation-drive}) is identical to that obtained for the one dimensional gas with driving [see Eq.~(21) of Ref.~\cite{Prasad:18}]. Hence, we can read out the results, which may be summarised as follows.
\be
\label{eq:casedriving}
\beta_d=\begin{cases}
\gamma,  &  r_w<1,\\
\min\left[\frac{2 + \min(\delta,0)}{2},\gamma \right], & r_w=1,
\end{cases}
\ee
where additional logarithmic corrections are present when $r_w=1$ and $\delta >0$. These logarithmic corrections take the form
\be
\ln P(\boldsymbol{v}) = - a |\boldsymbol{v}| (\ln |\boldsymbol{v}|)^{\frac{\gamma}{\gamma-1}} + \ldots,\quad r_w=1, \delta> 0, \gamma >1.
\label{pdf delta gt 0 rw1}
\ee
Thus, when $|r_w| <1$, the tails of the velocity distribution are similar to those of the noise distribution. When $r_w=1$, there is a regime where universal tails are seen. For $\delta \leq 0$, the exponent $\beta_d$ depends on $\delta$. For $\delta>0$ and $\gamma>1$, we see that the velocity distribution is an exponential ($\beta_d=1$) with additional logarithmic corrections. These logarithmic corrections are dependent on the noise characteristics~\cite{Prasad:18}.

We now focus on determining $\beta_c$ by looking for self consistent solutions of Eq.~(\ref{moment-equation-coll}). We note that for 
the one dimensional  gas $\beta_c=\infty$, and thus $\beta=\beta_d$~\cite{Prasad:18}. However, 
for the two dimensional gas,
it will turn out that  $\beta_c \neq \infty$, and there are regimes where $\beta_c<\beta_d$.

For large $n$, the summation
$\sum_m t_m$ in Eq.~(\ref{moment-equation-coll}) may be converted to an integral by changing variables to $y =m/( 2 n)$. We evaluate the integrals over $\theta$ and $y$ by the
saddle point approximations, valid for large $n$ (see \ref{appendix2} for details). 
For $\beta_c>2$, the maximum occurs for $0<y^* <1$. We then obtain for $\beta_c>2$:
\begin{align}
n^{\frac{\tilde{\delta}}{\beta_c}}M_{2n}&\sim \frac{n^{\frac{4+\delta}{\beta_c}}}{n^{5/2}} 
 \left[\frac{2n}{ae\beta_c}\right]^{\frac{2n}{\beta_c}}
\left[\frac{1}{1-y^*}\right]^{\frac{n (\beta_c-2)}{\beta_c}} 
 e^{\Psi[(\frac{2 n(1-y^*)}{a \beta_c})^{\frac{1}{\beta_c}}]
+\Psi[(\frac{2 n y^*}{a \beta_c})^{\frac{1}{\beta_c}}]}, 
\label{moment equation-no driving}\\
y^* &= \frac{\alpha^{\frac{\beta_c}{\beta_c-2}}}{\alpha^{\frac{\beta_c}{\beta_c-2}}+(2-\alpha)^{\frac{\beta_c}{\beta_c-2}}}.
\label{eq:ystar}
\end{align}
The result for $M_{2n}$ in  Eq.~\eref{moment equation-no driving}  is not consistent with the expression for $M_{2n}$ in Eq.~\eref{velocity-moment} due to 
the additional 
exponential term $(1-y^*)^{\frac{n (2-\beta_c)}{\beta_c}}$  in the right hand side of Eq.~\eref{moment equation-no driving}. The only way to compensate for this term is if the subleading correction $\Psi(x) \sim x^{\beta_c}$.
However, this contradicts our assumption that $\Psi(x) x^{-\beta_c} \to 0$.
Thus, we conclude that our assumption of $\beta_c>2$ must be incorrect and hence, we obtain the
bound:
\be
\beta_c \leq 2.
\label{eq:01}
\ee

For $\beta_c \le 2$, the maximal contribution from the integral comes from the endpoint $y=0$.
Then, the scaling $m=2 n y$ breaks down and it is possible that the maximal contribution to the
first sum in Eq.~(\ref{moment-equation-coll}) is  from a term with $m^*$ which scales with $n$ as $m^* \sim n^\phi$ with $\phi<1$. $m^*$ satisfies
$t_{m^*} \approx t_{m^*+1}$. To determine  $m^*$, we first
evaluate $t_m$ [see Eq.~\eref{moment equation-no driving-typical term}] in the limit $m\ll n$ to obtain
\begin{align}
\label{moment equation-no driving-typical term-theta-integrated}
t_{2m}\approx&\frac{(2n\alpha)^{2 m}}{(2m)!}n^{\frac{2+\delta-\beta_c}{\beta_c}}\left(\frac{2n}{ae\beta_c}\right)^\frac{2n}{\beta_c}\left(\frac{2n}{a\beta_c}\right)^\frac{-2m}{\beta_c} e^{\Psi[(\frac{2 n}{a \beta})^{\frac{1}{\beta}}]}\notag\\
\times&\displaystyle\sum_{j=0}^{m}\frac{\langle v_x^{2j} v_y^{2 m-2j}\rangle\Gamma(j+m+1/2)}{[\alpha(2-\alpha) n]^{m+j+1/2}}.
\end{align}
For large $n$, the sum in the right hand side of Eq.~\eref{moment equation-no driving-typical term-theta-integrated} is dominated by the term $j=0$, as every successive term is smaller by a factor of $n$. 
Taking the ratio of successive terms, we obtain
\be
\label{ratio t_k+1-by-t_k}
\frac{t_{2 m+2}}{t_{2 m}}\approx\frac{\alpha}{(2-\alpha)}\left(\frac{a\beta_c}{2}\right)^\frac{2}{\beta_c}\frac{\langle v_y^{2 m+2}\rangle}
{m \langle v_y^{2 m}\rangle}\frac{1}{n^{\frac{2-\beta_c}{2}}},\quad \frac{m}{n} \to 0.
\ee
When $\beta_c<2$, each successive term $t_m$ is smaller by a factor $n^{\frac{2-\beta_c}{2}}$. Thus, the largest term is $t_2$, 
and therefore, from Eq.~\eref{moment equation-no driving-typical term-theta-integrated}, we obtain
\be
\label{leading order t_1 betaless than 2}
n^{\frac{\tilde{\delta}}{\beta_c}}M_{2n} \sim t_2\sim n^{\frac{2+\delta-\beta_c}{\beta_c}+(\frac{1}{2}-\frac{2}{\beta_c})  }\left(\frac{2n}{ae\beta_c}\right)^\frac{2n}{\beta_c} e^{\Psi[(\frac{2 n}{a \beta})^{\frac{1}{\beta}}]}
\ee
Comparing with the expression for $M_{2n}$ in Eq.~\eref{velocity-moment}, we obtain 
 $1/2-2/\beta_c=(\tilde{\delta}-\delta)/\beta_c$ or $\beta_c=2 (2+ \tilde{\delta}-\delta)$. Since $\tilde{\delta}-\delta \geq 0$, we obtain $\beta_c \geq 4$ which contradicts our assumption that $\beta_c<2$. Thus,
$ \beta_c \geq 2$. This result, together with Eq.~\eref{eq:01}, implies that
\be
\beta_c=2.
\label{eq:02}
\ee
We now check when $\beta_c=2$ is a self-consistent
solution.

When $\beta_c=2$ the ratio of successive terms $t_m$, as in Eq.~\eref{ratio t_k+1-by-t_k}, simplifies to 
\be
\label{ratio t_k+1-by-t_kb2}
\frac{t_{2 m+2}}{t_{ 2m}}\approx\frac{\alpha a}{(2-\alpha)}\frac{\langle v_y^{2 m+2}\rangle}
{m \langle v_y^{2 m}\rangle},~\frac{m}{n} \to 0, ~\beta_c=2.
\ee
When $m^* \sim n^\phi$ with $0<\phi <1$, then the moments of the velocity may be evaluated using
Eq.~\eref{velocity-moment} to obtain $\langle v_y^{2 m+2}\rangle/
(m \langle v_y^{2 m}\rangle) = a^{-1}$, such that $t_{2 m+2}/t_{2 m} = \alpha/(2-\alpha)<1$. This implies
that successive terms are smaller, and therefore the solution for $m^*$ is such that $m^* \sim n^0$.
For any such $m^*$, it is straightforward
to obtain from Eq.~\eref{moment equation-no driving-typical term-theta-integrated} that
$M_{2n} \sim t_{m^*} \sim M_{2n}/\sqrt{n}$
which is not a consistent solution. 

We note that for $\beta_c>2$, the collision sum overestimates $M_{2n}$ [see Eq.~\eref{moment equation-no driving}]
while it underestimates $M_{2n}$ for $\beta_c\leq 2$. To obtain additional power law factors of $\mathcal{O}(\sqrt{n})$
for $\beta_c=2$, we
require $\langle v_y^{2 m}\rangle$ in Eq.~\eref{ratio t_k+1-by-t_k} to depend on $n$, such that $m^* \sim n^\phi$ with $\phi >0$. 
This is possible if
there are additional logarithmic corrections present in the velocity distribution such that
\be
P(\boldsymbol{v}) \sim \exp\left[-a|\boldsymbol{v}|^2 (\ln|\boldsymbol{v}|)^\tau\right],~~ |\boldsymbol{v}|^2 \gg \langle |\boldsymbol{v}|^2 \rangle.
\label{pdf-log}
\ee
For such a distribution, it may be shown that
\be
\label{ratio-log}
\frac{t_{2 m+2}}{t_{2m}}\approx\frac{\alpha}{2-\alpha}\left[\frac{\ln n}{\ln m} \right]^\tau ~~m\ll n.
\ee
Setting the ratio to $1$, we obtain $m^* \sim n^\phi$, where $\phi = [\alpha/(2-\alpha)]^{1/\tau}$.
Since $\phi<1$, we require $\tau>0$, such that the distribution decays faster than a
gaussian. Determining $\tau$ 
requires keeping more than the first few terms in the asymptotic behaviour of moments, which we are unable to currently do.

The exponent $\beta$ is now determined from Eqs.~(\ref{eq:equality}), (\ref{eq:casedriving}) and (\ref{eq:02}).
For $|r_w|<1$, a universal regime is reached if $\gamma >2$ in which case $\beta=\beta_c=2$. For $r_w=1$,
a universal regime is reached if $\gamma >\gamma^*(\delta)=[2+ \min(\delta,0)]/2$, in which case $\beta=\beta_d=\gamma^*(\delta)$. This may be summarised as
\be
\label{eq:casesolution}
\beta=\begin{cases}
\min[\gamma, 2] , &  r_w<1,\\
\min\left[\gamma, \frac{2 + \min(\delta,0)}{2} \right], & r_w=1.
\end{cases}
\ee

Our driving rules may be interpreted as  particles being driven through  collisions with a wall with $r_w$ being the coefficient of restitution between wall and particles. With such an interpretation, one would generically expect  $r_w<1$. Also, noise distributions
typically have a largest velocity, corresponding to large $\gamma$. This corresponds to the first case in Eq.~\eref{eq:casesolution}, and we 
conclude that the tails of the velocity distribution are generically a gaussian with
additional logarithmic corrections, as described in Eq.~\eref{eq:main}.

\section{\label{sec5-Monte-Carlo results}Monte Carlo results}

\begin{figure}
\begin{center}
 \includegraphics[width=1\textwidth]{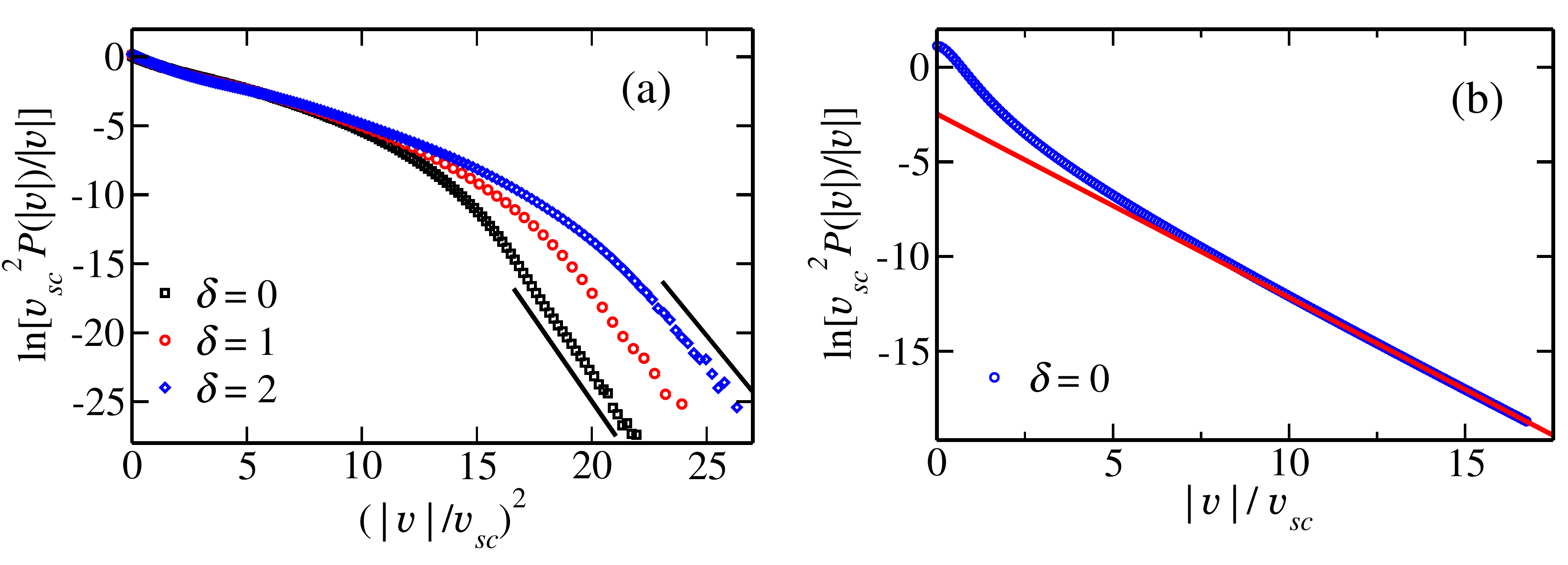}
 \end{center}
\caption{\label{MC-Simulation1}Results from Monte Carlo simulations for the scaled distribution $v^2_{sc} P(|{\boldsymbol v}|)/|{\boldsymbol v}|$ for a two dimensional inelastic gas for isotropic driving~[see~\eref{wall collision2} and~(\ref{ansatz noise pdf})], where $v_{sc}=v_{rms}/\sqrt{2}$.  (a)   The scaled distribution varies with  $({|\boldsymbol{v}}|/v_{sc})^2$ asymptotically as a straight line for  $\delta=0,1,2$, when  the noise  distribution $\Phi(|{\boldsymbol{\eta}}|)$ is  a uniform distribution, corresponding to $\gamma=\infty$.  (b)
The velocity distribution  decays exponentially (for  $\delta=0$), when  the noise  distribution $\Phi(|{\boldsymbol{\eta}}|)$ is exponential.  Notice that  here the scaled distribution is plotted against $(|{\boldsymbol{v}}|/v_{sc})$. The solid  straight line is guides to the eye.  The data are for $r_w=1/2$, $r=0$ and $\lambda_c=\lambda_d=1/2$.}
\end{figure}

In this section, we confirm that results from Monte Carlo simulations are consistent with our analytical results for the tail of the distribution.
We perform Monte Carlo simulation to obtain the steady state  distribution   
for the two dimensional inelastic gas with dissipative driving as described in Eq.~\eref{wall collision2}. All the simulations are for
$N=1000$, $r=0$ and $\lambda_c=\lambda_d=1/2$, and the data are averaged over the steady state.
 We first discuss the case when $r_w<1$. 
In \fref{MC-Simulation1}, we show
the variation of the scaled distribution of the modulus of  velocity,  $v^2_{sc} P(|{\boldsymbol v}|)/|{\boldsymbol v}|$~[$v_{sc}=v_{rms}/\sqrt{2}$, where $v_{rms}$ is the root mean square velocity] with scaled speed,  for  isotropic driving with $r_w=1/2$.
Figure~(\ref{MC-Simulation1})(a) shows the probability distribution for different values of $\delta=0,1,2$ when the noise distribution $\Phi({\boldsymbol{\eta}})$
is a uniform distribution in the range $|{\boldsymbol{\eta}}|<1$ corresponding to $\gamma=\infty$.  
When plotted against $({|\boldsymbol{v}}|/v_{sc})^2$, the linear behaviour for large velocities is consistent with
our prediction of $\beta=2$ for $\gamma >2$ [Eq.~\eref{eq:02}]. 
The tails are not sampled well enough to identify logarithmic corrections to this leading behaviour, if any.
In Figure (\ref{MC-Simulation1})(b) the velocity distribution is plotted for the Maxwell gas ($\delta=0)$, 
when the noise distribution $\Phi({\boldsymbol{\eta}})$ is an exponential, corresponding to  $\gamma=1$. 
When plotted against $({|\boldsymbol{v}}|/v_{sc})$, the linear behaviour for large velocities
is consistent with our prediction of $\beta=\gamma$ for $\gamma \leq 2$.
Thus, the results from Monte Carlo simulations with isotropic driving are consistent with the analytical results that we have obtained  for $\beta$.

\begin{figure}
\begin{center}
 \includegraphics[width=1\textwidth]{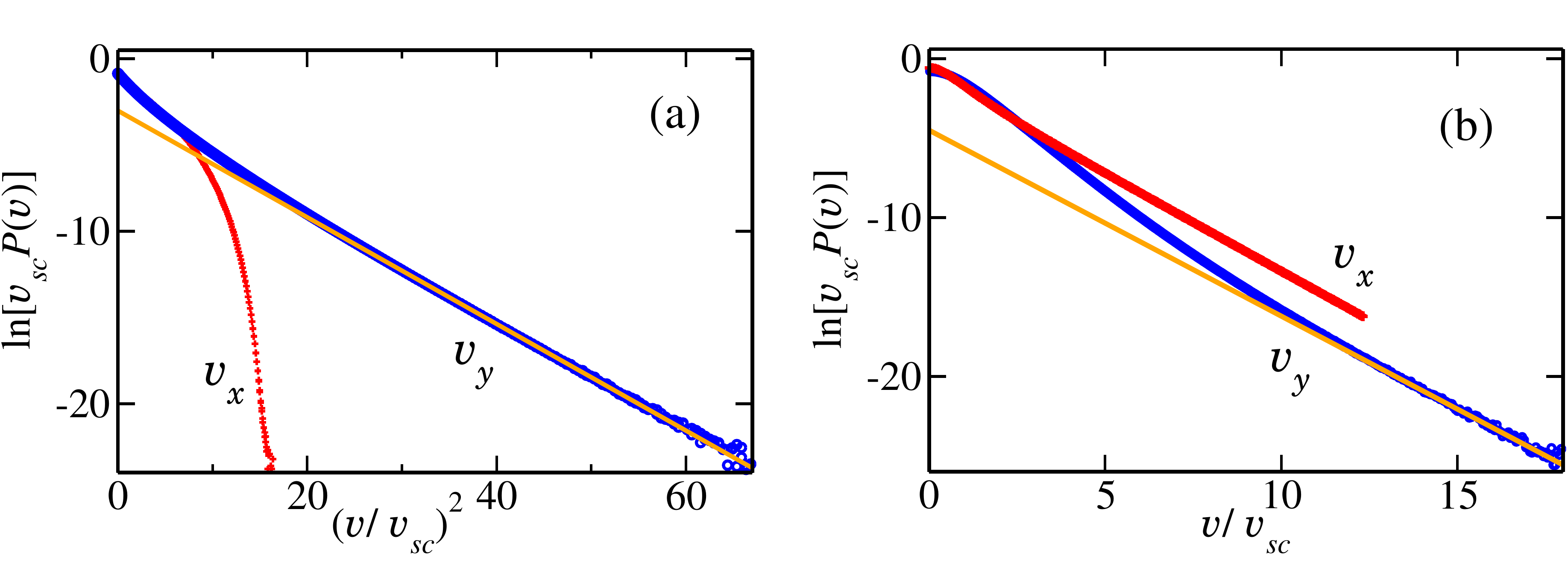}
 \end{center}
\caption{\label{Fig-anisotropic}(Color Online) The steady state velocity distribution ($x$ and $y$ components) of a two-dimensional Maxwell gas 
when only the $x$-component is driven, as obtained from Monte Carlo simulations. Here, $v_{sc}= \sqrt{\langle v_{x,y}^2 \rangle}$. The data are for when the noise distribution $\Phi(\eta_x)$  is uniform  in [-3/2,3/2]~[plot~(a)], and an exponential~[plot~(b)]. 
The solid  straight lines are guides to the eye.}
\end{figure}

We also  present results from Monte Carlo simulations for the steady state velocity distribution,
when driving is restricted to the
$x$-component, mimicking many 
experiments~\cite{Olafsen:98,Olafsen:99,Losert:99,Kudrolli:00,Rouyer:00,Blair:01,Vanzonexpt:04} where particles are driven in one direction
and  the distribution of the velocity component perpendicular to the driving is measured.
Figure~\ref{Fig-anisotropic} shows the results for the scaled distribution of $x$ and $y$ components of velocities, $v_{sc}P(v)$~[$v_{sc}=$ root mean square velocities of $v_x,~v_y$ respectively] of a Maxwell gas~[$\delta=0$] when only the $x$-component of the velocity is driven and the noise distribution $\Phi(\eta_x)$ is 
chosen to be uniform ($\gamma=\infty$) [see \fref{Fig-anisotropic}(a)] and
an exponential ($\gamma=1$) [see \fref{Fig-anisotropic}(b)]. When the noise distribution is uniform,  $P(v_y)$ is consistent with a Gaussian with $\beta=2$. When the noise distribution is an exponential, $P(v_y)$ is consistent
with an exponential distribution. Thus, the results for the non-driven component are  as predicted by our calculation for the isotropic problem, showing that the analytical results possibly extend to anisotropic driving also.

\begin{figure}
\begin{center}
 \includegraphics[width=1\textwidth]{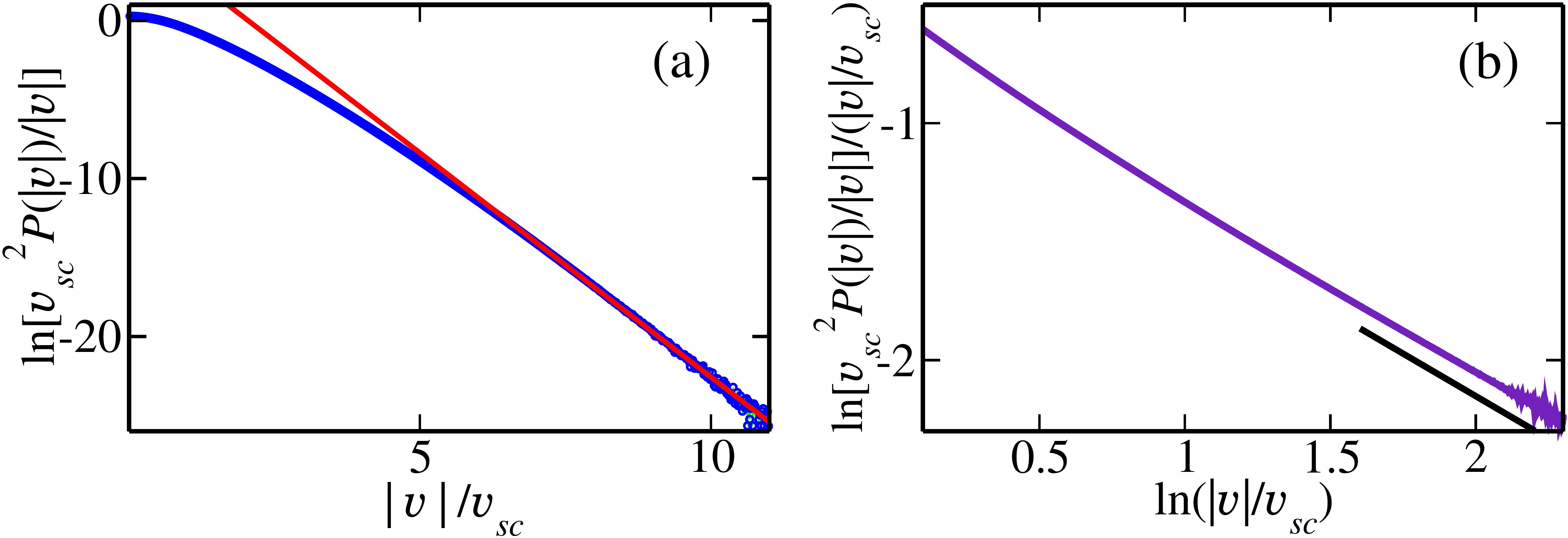}
 \end{center}
\caption{\label{Fig-rw1-delta1}(Color Online) The steady state velocity distribution of a two-dimensional granular gas 
($\delta=1$) for $r_w=1$  as obtained from Monte Carlo simulations for a isotropically driven system when the noise distribution is a  uniform distribution. (a) The data is a straight line near the tails when plotted against scaled speeds. (b) The same data as in (a) but when the distribution is divided by speed, in order to obtain subleading corrections (Note that the x-axis is  now $\ln[{|\boldsymbol{v}}|/v_{sc}]$, where $v_{sc}=v_{rms}/\sqrt{2}$). The solid  straight lines are guides to the eye.}
\end{figure}

We now focus on the case $r_w=1$. This special case corresponds to diffusive driving in kinetic theory. We also focus on the case $\delta=1$, when the collisions between particles are proportional to the relative velocity. Our analytical results predict that the velocity distribution is an exponential with logarithmic corrections. We now show that the numerical results are consistent with this prediction.
In \fref{Fig-rw1-delta1}(a), we show
the variation of the scaled distribution of the modulus of  velocity,  $v_{sc}^2 P(|{\boldsymbol v}|)/|{\boldsymbol v}|$ with scaled speed, for  isotropic driving with $r_w=1$, when the noise distribution is uniform distribution in the range $|{\boldsymbol{\eta}}|<1$ corresponding to $\gamma=\infty$.  
When plotted against ${|\boldsymbol{v}}|/v_{sc}$, the linear behaviour for large velocities is consistent with
our prediction of $\beta=1$ for $\gamma >1$ [Eq.~\eref{eq:main2}]. We now check whether the logarithmic corrections to the exponential distribution is captured by the simulations. 
In \fref{Fig-rw1-delta1}(b), we show the same data as in \fref{Fig-rw1-delta1}(a), but after dividing
$\ln[v_{sc}^2 P(|{\boldsymbol v}|)/|{\boldsymbol v}|]$  by the scaled speed.
When plotted against $\ln[{|\boldsymbol{v}}|/v_{sc}]$, the linear behaviour for large velocities is consistent with
our prediction of $\theta=1$ for $\gamma=\infty$ [Eq.~\eref{eq:02}].
Thus, the results from Monte Carlo simulations with isotropic driving for $r_w=1$ are consistent with the analytical results that we have obtained  for $\beta$.

\begin{figure}
\begin{center}
 \includegraphics[width=0.6\textwidth]{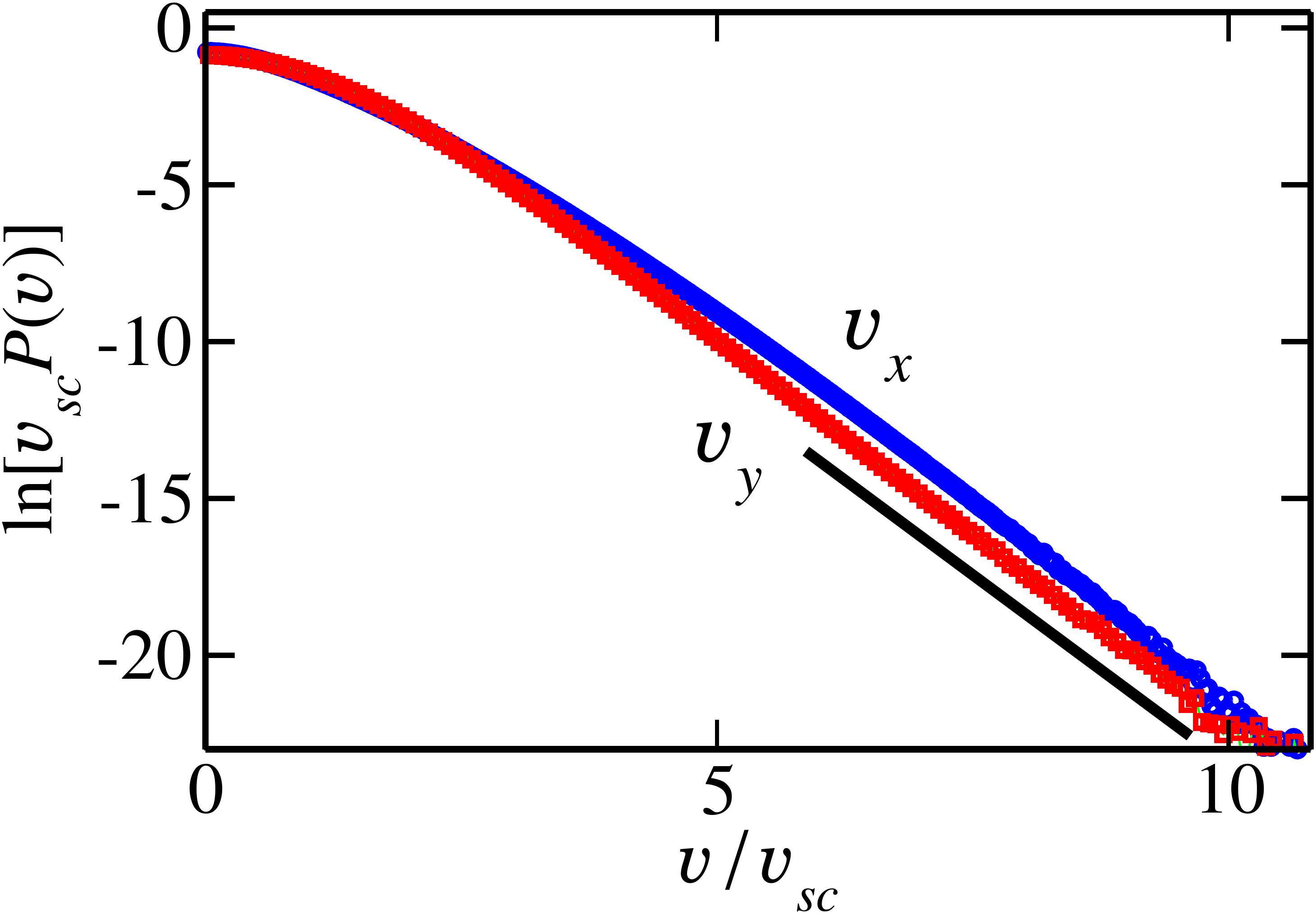}
 \end{center}
\caption{\label{Fig-rw1-delta1-anisotropic}(Color Online) The steady state velocity distribution of a two-dimensional granular gas 
($\delta=1$) for $r_w=1$  as obtained from Monte Carlo simulations when only the $x$-component is driven. The noise distribution is a uniform distribution. The data for $P(v_y)$ is a straight line near the tails when plotted against scaled speeds, where $v_{sc}= \sqrt{\langle v_{x,y}^2 \rangle}$. The solid  straight line is a guide to the eye.}
\end{figure}

We now confirm that, for $r_w=1$ also, the results do not change if the driving is anisotropic and only one component is driven. In \fref{Fig-rw1-delta1-anisotropic}, we show
the variation of the scaled distribution of the velocities of $x$ and $y$ components,  $v_{sc} P( v)$ with scaled velocity $v/v_{sc}$~[$v_{sc}=$   root mean square velocity of the $v_x, v_y$ respectively], for anistropic driving with $r_w=1$, when the noise distribution is uniform distribution in the range $|\eta_x|<1/2$ corresponding to $\gamma=\infty$.  
The linear behaviour of $P(v_y)$ for large velocities is consistent with
our prediction of $\beta=1$ for $\gamma >1$ [Eq.~\eref{eq:main2}]. Thus, the results for the non-driven component are  as predicted by our calculation for the isotropic problem, showing that the analytical results possibly extend to anisotropic driving also.

\section{\label{sec7-Comparison with exp data}Comparison with previous experimental data} 

\begin{figure}
\begin{center}
 \includegraphics[width=1\textwidth]{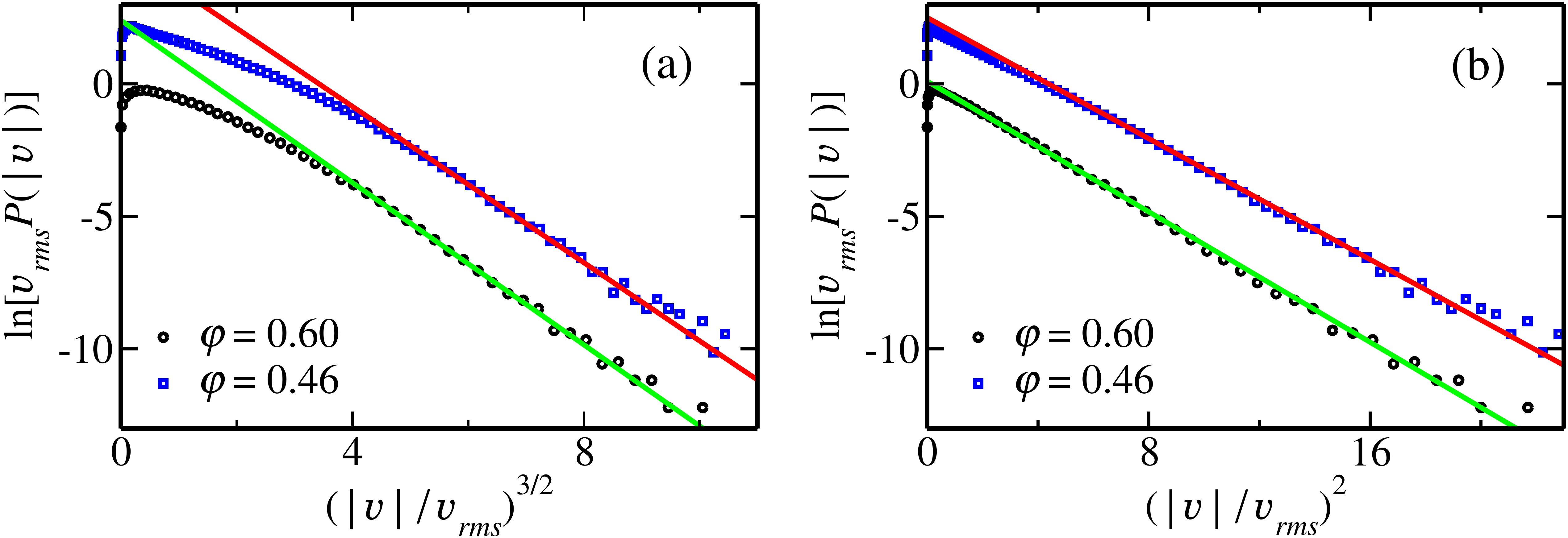}
 \end{center}
\caption{\label{poschel together}Experimental data for the velocity distribution, extracted from Fig.~5 of Ref~\cite{Scholz:17}, is plotted as a function of
$(|\boldsymbol{v}|/v_{rms})^{3/2}$~[plot (a)] and $(|\boldsymbol{v}|/v_{rms})^{2}$~[plot (b)]. The data are for two different volume fractions $\varphi$. As in Ref.~\cite{Scholz:17} one of the data-set is shifted vertically  for clarity. Solid straight lines are guides for the eye. 
}
\end{figure}

We note that the experimental data for the measurement of $P(\boldsymbol{v})$ may be open to interpretation. As an example,
by re-plotting, we show that the data obtained in a recent experiment~\cite{Scholz:17} with homogeneous driving, and has been argued for evidence for $\beta=3/2$, are also consistent with $\beta=2$.
The experiment involved a  system of particles  residing on a two-dimensional surface which is driven through a periodic motion of the
surface so that the system is homogeneously driven. In addition, the rotational degrees of freedom are driven through collisions with the wall, which in turn drives the translational degrees of freedom. 
We extract the data for the velocity distribution from Fig.~5  of Ref.~\cite{Scholz:17}
and plot the data both as a function of the $(|\boldsymbol{v}|/v_{rms})^{3/2}$ and as a function of $(|\boldsymbol{v}|/v_{rms})^2$ [see \fref{poschel together}]. Clearly, the data cannot be used to distinguish between the two distributions. If anything, the Gaussian describes
the data better. The experiment is clearly more complicated than our model, where rotational degrees of freedom are ignored. But our simulations would suggest that the results are not sensitive to how the system is driven, rather it depends only on whether there is dissipation when particles collide with the wall. In this case, we expect the experiment to fall into the category of $r_w<1$, and hence $\beta=2$.

\section{\label{sec8-Discussions}Summary and discussion}

To summarise, in this paper, we considered a particle-based microscopic model for a driven granular gas in two dimensions. Energy is pumped into the system by driving the particles at a constant rate in a homogeneous, isotropic fashion using the driving rule described in Eq.~(\ref{wall collision2}). At every instance of driving, the velocity of the particle is reduced by a factor $r_w$ along with an additive noise chosen from a fixed distribution. For $r_w\neq -1$, the system reaches a steady state.  The rate of collision for a pair of particles is proportional to $|\boldsymbol{v}_{\rm{rel}}|^\delta$  where $\boldsymbol{v}_{\rm{rel}}$ is the relative velocity between them. In the well mixed limit, when the spatial correlations may be ignored, we determine analytically the tail of the velocity distribution by analysing in detail the equations satisfied by the moments of the velocity. This is done by assuming a certain stretched exponential form for the tails of the velocity distribution and finding self-consistent solutions.

\begin{figure}
\begin{center}
 \includegraphics[width=1\textwidth]{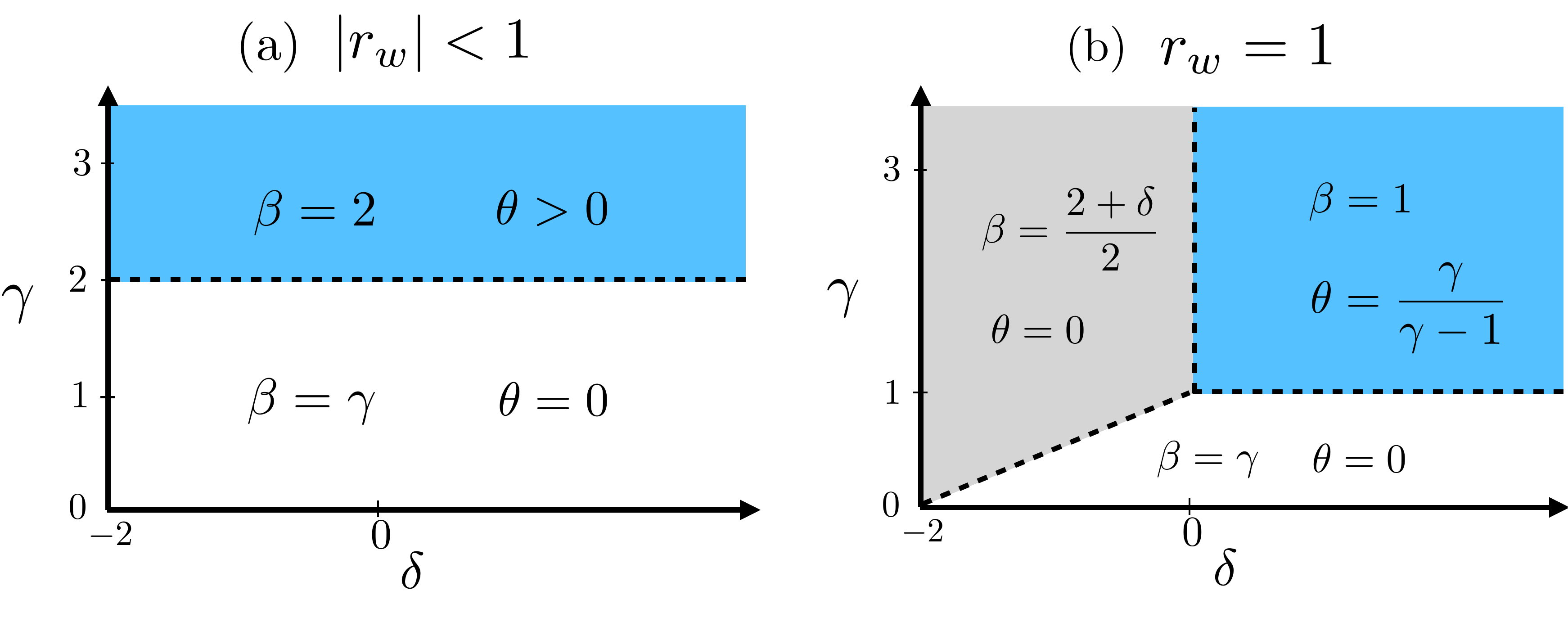}
 \end{center}
\caption{\label{Fig-summary}(Color Online) Summary of results for the stretched exponential exponent $\beta$  as a function of the parameters 
$\gamma$ and $\delta$, where $\ln P(\boldsymbol{v}) \sim -a |\boldsymbol{v}|^\beta (\ln |\boldsymbol{v}|)^\theta$. (a) $r_w\neq1$, (b)   $r_w=1$.  The shaded regions correspond to universal regimes where the tails of $P(\boldsymbol{v})$ are largely independent of the noise distribution.}
\end{figure}

Our main results are summarised in Eqs.~\eref{eq:main}-\eref{eq:main3}. The results depend on whether $|r_w|<1$ or $r_w=1$. When  $|r_w|<1$, there is a universal regime when the noise distribution decays faster than a gaussian ($\gamma >2$), and a non-universal regime when $\gamma \leq 2$. In the universal regime, the velocity distribution decays as a gaussian with logarithmic corrections. In the non-universal regime, $\beta=\gamma$. These results are summarised in \fref{Fig-summary}(a). Note that the results do not depend on $\delta$. When $r_w=1$, again we find a universal regime and a non-universal regime. Now, the results depend on $\delta$, as summarised in \fref{Fig-summary}(b). In the universal regime $\beta=1$, and in the non-universal regime $\beta=\gamma$. Note that $r_w=1$ corresponds to the scenario described by kinetic theory, and for $\delta=1$, we obtain $\beta=1$ in contradiction to the kinetic theory result of $3/2$. We note that $\gamma=0$  includes noise distributions that are power law distributed and $\gamma=\infty$ includes noise distributions that decay faster than stretched exponentials. 

The origin of these two universal regimes are quite different. For $|r_w|<1$, it is the collision between particles that results in a universal tail. This mechanism crucially depends on energy transfer between $x$ and $y$-components through collisions which are not head on. This mechanism is absent for one dimensional gases, and hence this universal regime is absent for such gases~\cite{Prasad:18}. For models in higher dimensions, energy transfer between the components of velocities is still possible, and we expect universal tails. For $r_w=1$, the universal regime is achieved through a balance between generation of large speeds due to driving and loss of large speeds due to collision.

We note that while microscopic models for  driven inelastic gases 
have been explored earlier, it has been difficult to obtain results for $\delta \neq 0$, when the rate of collisions depend on the relative velocity. Analysis of moment equations allows us to determine results for $\delta \neq 0$. Also see Ref.~\cite{Prasad:18}, where this could be accomplished for a one-dimensional granular gas.

 All the analytical results derived in this paper were based on dissipative driving and  homogeneous, isotropic noise. One of the motivations for dissipative driving is that in scenarios when the particles are driven through collisions with a wall,  the wall-collisions are inherently dissipative.   This analogy  is strictly valid only in one dimension, as it assumes that the wall moves colinearly with the particle velocity. For a two dimensional gas, one would expect that only the component of velocity perpendicular to the motion of the wall is reversed. A realistic model would be one where only one of the two components of velocity is reversed when driven, making the noise anisotropic. The generalisations of the calculations to this case is nontrivial. However, we have performed Monte Carlo simulations of anisotropic driving and found that the results for the asymptotic behaviour of the velocity distribution is in agreement with what we found analytically for isotopic driving. This is a strong reason to believe that our results are applicable to more general situations than the case for isotropic noise.

Our results are consistent with experimental~\cite{Baxter:03,Falcon:2013} studies as well.
It was shown that gaussian tails exist for the velocity distribution in realistic systems~\cite{Baxter:03}
when driven mechanically. We have also shown that data in Ref.~\cite{Scholz:17} may be re-interpreted as being a gaussian. 
By contrast,  experiment on magnetically driven granular 
matter~\cite{Falcon:2013} has illustrated tails that decay exponentially. We believe that the latter could be
explained by our analytical result for $r_w=1$. In the experiment on magnetically driven systems, 
driving affects only the rotational degrees of freedom, and could be compared with our driving model
with $r_w=1$, which changes the sign of the velocity components without dissipating energy.

We now discuss why kinetic theory result ($\beta=3/2$) differs from our results.
In kinetic theory, the steady state one particle scaled 
velocity distribution $P(v)$ satisfies~\cite{Noije:98}: 
\be
-c_1 v^{\delta} P(v) + c_2 \frac{\partial^2}{\partial v^2} P(v) = 0,  
\label{diff_drive}
\ee
where the two  terms describe collisional losses and  gains due to driving, where the gains due to interparticle collisions have been dropped. Assuming
$P(v) \sim \exp(-a v^{\beta})$, it is easy to show from Eq.~\eref{diff_drive} that $\beta=(2+\delta)/2$; for 
$\delta=1$ (in the ballistic case), $\beta = 3/2$~\cite{Noije:98}. The diffusive driving term in Eq.~\eref{diff_drive} may be derived
from our model when $r_w=1$ and Taylor-expanding for small $\eta$. However,
it may be shown that truncation at $\mathcal{O}(\eta^2)$  is not in general valid for $\delta>0$ when considering the tails of the distribution, and that
the largest term in the Taylor expansion corresponds  to a higher order derivative. This analysis may be found in Appendix B of Ref.~\cite{Prasad:18}. 

Boltzmann equation with a dissipative term,
\be
-c_1 v^{\delta} P(v) + c_2 \frac{\partial^2}{\partial v^2} P(v)+ \Gamma \frac{\partial}{\partial v}[v P(v)] = 0,
\label{dissi_drive}
\ee
has also been studied where the origin of the dissipation is due to either near-elastic wall collisions~\cite{Prasad:14}
or thermostatting~\cite{Montanero:00,Biben02}. By balancing the second and third terms and ignoring
inter-particle collisions, one obtains $\beta=2$.
However, neither does it capture the logarithmic corrections, not does it capture the correct physics, as the
origin of $\beta =2$ in our calculation is  inter-particle collisions.

In the present study we have assumed the system to be  well-mixed. This allows one to integrate out
spatial degrees and consider the system in a mean-field setup.  It would be important to check numerically,
how spatial dependence affect the results obtained here.  Determining these in 
large scale simulations in two and three dimensions are promising areas for future study. 

Also, for $r_w<1$, we have not been able to determine analytically
the logarithmic corrections that are present in the tails of the velocity distribution, neither are we able to unambiguously measure the logarithmic corrections in the Monte Carlo simulations. Monte Carlo simulations which are biased towards rare events through importance sampling could be one way of addressing this drawback. We are currently working on this problem.

\section*{Acknowledgments}
This research was supported in part by the International Centre for Theoretical Sciences (ICTS) during a 
visit for participating in the program -Indian Statistical Physics Community Meeting 2016 (Code: ICTS/Prog-ISPC/2016/02)

\appendix

\section{Large Moments--Asymptotic behaviour \label{appendix1}}

In the following, we derive the asymptotic form for the moments of the velocity distribution for large $n$. 
Consider an isotropic  velocity distribution having the form:
\begin{align}
P(\boldsymbol{v}) \sim e^{-a|\boldsymbol{v}|^\beta+\Psi(|\boldsymbol{v}|)},~a,\beta>0,|\boldsymbol{v}|^2 \gg \langle |\boldsymbol{v}|^2 \rangle,
\label{SI-velocity-isotropic pdf}
\end{align}  
and  $\Psi(|\boldsymbol{v}|)/|\boldsymbol{v}|^\beta\to 0$.
We would like to compute the moments of $(a_i v_x+a_j v_y)^{2 n}$, for large $n$ . This may be formally written as
\begin{align}
\langle (a_i v_x+a_j v_y)^{2 n}\rangle\sim\displaystyle\int dv_x dv_y  (a_iv_x+a_jv_y)^{2n}  e^{a (v_x^2+v_y^2 )^\frac{\beta}{2}+\Psi([v_x^2+v_y^2]^{\frac{1}{2}})}
\label{SI-velocity-isotropic-moment}
\end{align} 
Changing variables to  $v_x=n^{1/\beta }t$ and $v_y=n^{1/\beta}u$  the integral in \eref{SI-velocity-isotropic-moment} may be rewritten 
as
\begin{align}
\langle(a_i v_x+a_j v_y)^{2 n}\rangle\sim n^\frac{2}{\beta}n^\frac{2n}{\beta}   \displaystyle\int dt~ du e^{nf(t,u)}e^{\Psi\left(n^{\frac{1}{\beta}}\left[t^2+u^2\right]^{\frac{1}{2}}\right)},
\label{SI-velocity-isotropic-moment-new-variable}
\end{align}
where 
\be
f(t,u)=2 \ln[a_it +a_j u]-a (t^2+u^2)^\frac{\beta}{2}.
\ee
Doing a saddle point integration by maximising $f(t,u)$ with respect to the 
variables $u$ and $t$, we obtain:
\begin{align}
\langle(a_i v_x+a_j v_y)^{2 n}\rangle\sim n^\frac{2+2n}{\beta} e^{\Psi\left(n^{\frac{1}{\beta}}\left[{t^*}^2+{u^*}^2\right]^{\frac{1}{2}}\right)}\frac{e^{nf(t^*,u^*)}}{n} ,~~n\gg1,
\label{SI velocity-isotropic-moment-new-variable-saddle int}
\end{align}
where the integration over $u$ and $t$ pulls down a factor $1/n$.  $(t^*,u^*)$ is the point at which
$f$ has a maximum,
\begin{align}
t^*&=\frac{a_i}{\sqrt{a_i^2+a_j^2}}\left({\frac{2}{a\beta}}\right)^\frac{1}{\beta},\\
u^*&=\frac{a_j}{\sqrt{a_i^2+a_j^2}}\left({\frac{2}{a\beta}}\right)^\frac{1}{\beta}.
\label{SI max values s t}
\end{align}
Substituting for $t^*$ and $u^*$  in \eref{SI velocity-isotropic-moment-new-variable-saddle int}, we obtain
\begin{align}
\langle(a_i v_x+a_j v_y)^{2 n}\rangle\sim \left(\frac{2n}{ae\beta}\right)^{\frac{2n}{\beta}}  n^{\frac{2}{\beta}-1} e^{\Psi\left(\left[\frac{2n}{a\beta}\right]^{\frac{1}{\beta}}\right)} \left(a_i^2+a_j^2\right)^n, ~~n\gg1,
\label{SI velocity-isotropic-moment-new-variable-saddle int-final1}
\end{align}
and in particular, when $a_i=1$ and $a_j=0$, the result simplifies to
\begin{align}
M_{2n}=\langle v_x^{2 n}\rangle\sim \left(\frac{2n}{ae\beta}\right)^{\frac{2n}{\beta}} n^{\frac{2}{\beta}-1}e^{\Psi\left(\left[\frac{2n}{a\beta}\right]^{\frac{1}{\beta}}\right)},~~n\gg1.
\label{SI velocity-isotropic-moment-new-variable-saddle int-final2}
\end{align}

 Now consider the expression
\be
\langle |\boldsymbol{v}-\boldsymbol{v}'|^\delta
(a_1 v_x+a_2 v_y)^{2 n-2 m} (a_3 v'_x+a_4 v'_y)^{2 m}\rangle. \notag
\label{velocity-coupled-moment}
\ee
When $n, m \gg 1$, then on scaling the velocity components by $n^{1/\beta}$ as in Eq.~(\ref{SI-velocity-isotropic-moment-new-variable}), the factor $ |\boldsymbol{v}-\boldsymbol{v}'|^\delta$ yields a factor $n^{\delta/\beta}$. The integrals then decouple for large $m$ and $n$, and using Eq.~(\ref{SI velocity-isotropic-moment-new-variable-saddle int-final1}), we obtain
\be
\langle |\boldsymbol{v}-\boldsymbol{v}'|^\delta
(a_1 v_x+a_2 v_y)^{2 n-2 m} (a_3 v'_x+a_4 v'_y)^{2 m}\rangle \sim n^{\frac{\delta}{\beta}} M_{2 n -2 m} M_{2m} (a_1^2 \!+\! a_2^2)^{n-m} (a_3^2 \!+ \!a_4^2)^{m}.
\label{eq:see1}
\ee

\section{Collision sum: Asymptotic expression \label{appendix2}}

 In this section we consider the collision contribution in the moment equation~[Eq.~(\ref{moment-equation-coll}) in main text], 
\begin{align}
I=\displaystyle\sum_{m=0}^{2n } \binom{2n}{m}  \int_0^{2 \pi}  \frac{d\theta}{2 \pi}   \langle |\boldsymbol{v}-\boldsymbol{v}'|^\delta
(a_1 v_x+a_2 v_y)^{2 n-m} (a_3 v'_x+a_4 v'_y)^{m}\rangle,
\label{SI collision sum}
\end{align}
with
\begin{align}
\begin{split}
a_1=1-\alpha\cos^2\theta,~~&a_2=-\alpha\cos\theta\sin\theta\\ 
a_3=1-a_1, ~\text{and} ~&~~a_4=-a_2.
\label{SI-angle-a-s}
\end{split}
\end{align}
Using the asymptotic expression for the moments [Eqs.~\eref{SI velocity-isotropic-moment-new-variable-saddle int-final1}-\eref{eq:see1}],
 it may be seen that each of the terms
in the sum grows factorially with $n$ and therefore we perform the sum by the saddle point approximation. Taking Stirling's approximation,  substituting the values of $a_i$'s and  converting the sum to integral with respect to a variable  $y=m/(2 n)$ Eq.~\eref{SI collision sum} reduces to
\begin{align}
I\sim n^{\frac{4+\delta}{\beta}-\frac{3}{2}}\int_0^1dy\int_0^{2 \pi} \!\! \frac{d\theta}{2 \pi}
[y(1-y)]^{\frac{2}{\beta}-\frac{3}{2}} e^{\Psi\left(\left[\frac{2n(1-y)}{a\beta}\right]^{\frac{1}{\beta}}\right)} e^{\Psi\left(\left[\frac{2n(y)}{a\beta}\right]^{\frac{1}{\beta}}\right)}e^{nf(\theta,y)},
\label{SI collision int}
\end{align}
where
\begin{align}
f(\theta,y)&=\frac{2(1-\beta)}{\beta}\left[y\ln y+(1-y)\ln(1-y)\right]+(1-y)\ln[1-\alpha(2-\alpha)\cos^2\theta]\nonumber \\
&+y\ln(\alpha^2\cos^2\theta)
\end{align}

The function $f(\theta,y)$ is maximised  when $\theta=\theta_{1,2}^*$ and $y=y^*$. It is easily obtained that
\begin{align}
\theta_1^*&=0,~\pi, \\
\cos^2 \theta_2^*&=\frac{y_2^*}{\alpha(2-\alpha)}.
\end{align}
For $\theta=\theta_1^*$ we obtain
\begin{align}
y_1^*&=\alpha^{\frac{\beta}{\beta-1}}\left[\alpha^{\frac{\beta}{\beta-1}}+(1-\alpha)^{\frac{\beta}{\beta-1}}\right]^{-1}, \\
f(\theta_1^*,y^*_1)&=\frac{2(\beta-1)}{\beta}\ln\left[\alpha^{\frac{\beta}{\beta-1}}+(1-\alpha)^{\frac{\beta}{\beta-1}} \right],
\end{align}
and for $\theta=\theta_2^*$, 
\begin{align}
y_2^*&=\alpha^{\frac{\beta}{\beta-2}}\left[\alpha^{\frac{\beta}{\beta-2}}+(2-\alpha)^{\frac{\beta}{\beta-2}}\right]^{-1}, ~\cos^2 \theta_2^*=\frac{y_2^*}{\alpha(2-\alpha)},~\\
f(\theta_2^*,y^*_2)&=\frac{(\beta-2)}{\beta}\ln\left[\frac{\alpha^{\frac{\beta}{\beta-2}}+(2-\alpha)^{\frac{\beta}{\beta-2}} }{ (2-\alpha)^{\frac{\beta}{\beta-2}} }\right].
\end{align}
A straightforward analysis of the two solutions may be performed to check which has a larger value for $f$. 
Specifically,  when $\beta>2$ the $f(\theta,y)$ has a maximum at $(\theta_2^*, y_2^*)$, but when $\beta\leq2$ it is at 
(0,0).

\section*{References}
\bibliographystyle{iopart-num}
\providecommand{\newblock}{}

\end{document}